\documentclass[number,preprint,3p]{elsarticle}

\usepackage{multirow}
\usepackage{color}
\usepackage{graphicx}
\usepackage{subcaption}
\usepackage{algorithm}
\usepackage{bm}
\usepackage[colorlinks]{hyperref} 
\usepackage{amssymb}
\usepackage{amsthm}
\usepackage{amsmath}
\usepackage{booktabs}
\usepackage{url}
\usepackage{scrextend}
\usepackage{epstopdf}
\usepackage{float}
\usepackage{tablefootnote}
\usepackage[table]{xcolor} 
\usepackage{mathtools}
\usepackage{soul}
\usepackage[perpage]{footmisc} 

\usepackage{lineno}

\makeatletter
\gdef\urlauthor#1#2{\g@addto@macro\@elsuads{\let\corref\@gobble%
     \def\@@tmp{#1}\raggedright\eadsep
     {\ttfamily\url{\expandafter\strip@prefix\meaning\@@tmp}}\space(#2)%
     \def\eadsep{\unskip,\space}}%
}
\gdef\emailauthor#1#2{\stepcounter{ead}%
     \g@addto@macro\@elseads{\raggedright%
      \let\corref\@gobble\def\@@tmp{#1}%
      \eadsep{\ttfamily\href{mailto:\expandafter\strip@prefix\meaning\@@tmp}{\expandafter\strip@prefix\meaning\@@tmp}}
      (#2)\def\eadsep{\unskip,\space}}%
}
\makeatother

\newcommand{\vect}[1]{\boldsymbol{#1}}

\makeatletter
\let\@afterindenttrue\@afterindentfalse
\makeatother

\biboptions{numbers,sort&compress,square}
\journal{arXiv}
\begin{document}
\begin{frontmatter}
\renewcommand{\thefootnote}{\fnsymbol{footnote}}

\title{Recorded Versus Synthetic Spectral-compatible Ground Motions: A Comparative Analysis of Structural Seismic Responses}

 \author[1]{Jungho Kim}
 \author[2]{Maijia Su}
 \author[3]{Ziqi Wang\corref{cor1}}
\ead{ziqiwang@berkeley.edu}  
         \cortext[cor1]{Corresponding author}
 \author[2]{Marco Broccardo\corref{cor2}}
 \ead{marco.broccardo@unitn.it}  
         \cortext[cor2]{Corresponding author}
 \address[1]{Department of Civil and Environmental Engineering, Sejong University, Seoul, South Korea}
 \address[2]{Department of Civil, Environmental and Mechanical Engineering, University of Trento, Italy}
 \address[3]{Department of Civil and Environmental Engineering, University of California, Berkeley, United States}
\begin{abstract}
This paper presents a comparative analysis of structural seismic responses under two types of ground motion inputs: (i) synthetic motions generated by stochastic spectral-compatible ground motion models and (ii) recorded motions from an earthquake database. Both ground motion datasets are calibrated to a shared target response spectrum to ensure consistent spectral median, variance, and correlation structure. Five key stochastic response metrics—probability distributions, statistical moments, correlations, tail indices, and variance-based global sensitivity indices—are systematically evaluated for two representative structures: a medium-period building and a limiting case of a long-period tower. The comparison accounts for uncertainties both from ground motion and structural parameters. The results reveal that synthetic motions closely replicate recorded motions in terms of global response behavior—including distributions, mean and variance, correlation structure, and dominant uncertainty sources—indicating their suitability for routine seismic design and parametric studies. However, substantial differences emerge in response extremes for long-period structures, particularly in metrics governed by rare events, such as higher-order moments and tail behavior. These differences, which often exceed 50\%, can be  attributed to the non-Gaussian features and complex characteristics inherent in recorded motions, which are less pronounced in synthetic datasets. The findings support the use of synthetic ground motions for evaluating global seismic response characteristics, while highlighting their limitations in capturing rare-event behavior and long-period structural dynamics.

\end{abstract}

\begin{keyword}
Recorded ground motion \sep stochastic seismic response \sep synthetic ground motion \sep variance-based global sensitivity analysis

\end{keyword}
\end{frontmatter}
\renewcommand{\thefootnote}{\fnsymbol{footnote}}

\section{Introduction}

\noindent Quantifying the variability in seismic responses of structures subjected to ground motion inputs is a fundamental challenge in performance-based earthquake engineering (PBEE) \cite{moehle2004framework,broccardo2016application,iervolino2017assessing,o2019conceptual,zhong2023surrogate} and seismic risk assessment \cite{ellingwood2009quantifying,jalayer2010structural,taflanidis2011simulation,kim2025efficient}. Structural responses to seismic excitations are primarily governed by two sources of input uncertainty: variability in ground motion characteristics and uncertainties in structural parameters. A comprehensive understanding of stochastic structural responses is crucial for enhancing the reliability of structural designs and accurately assessing seismic risks, particularly under extreme scenarios involving limit state exceedance. Evaluating different seismic response metrics enables systematic comparison of uncertainties across diverse scenarios, supporting risk-informed structural design.

The current practice for defining ground motion inputs typically involves generating a target response spectrum to characterize the seismic hazard at a site. The uniform hazard spectrum (UHS), derived from probabilistic seismic hazard analysis, provides spectral accelerations at various periods corresponding to a constant exceedance probability (e.g., 2\% probability of exceedance in 50 years). However, UHS does not represent the response spectrum of any single ground motion and is generally conservative. This limitation has led to the development of advanced spectra that incorporate additional probabilistic information, such as spectral acceleration variability \cite{wang_ground_2011} and spectral correlation across distinct periods \cite{baker2008correlation}. An advanced spectrum containing a complete probability distribution is termed an ``unconditional spectrum'' \cite{baker2018improved}. When response spectra follow a joint log-normal distribution, the unconditional spectrum is fully defined by the mean, variance, and correlation of spectral accelerations across a specific range of periods. Using the properties of the joint lognormal distribution, Baker \cite{baker2011conditional} developed the conditional mean spectrum (CMS), which refines spectral shapes by conditioning spectral accelerations at multiple periods on a target period of interest \cite{baker2011conditional}.

Provided with a target response spectrum, PBEE often requires selecting or generating spectrum-consistent ground motions. Ground motions are typically categorized into two types: recorded motions obtained from past earthquake events and synthetically generated ground motions. 
Recorded motions are widely regarded as realistic representations of earthquake phenomena, making them a preferred choice in design contexts due to their direct connection to observed seismic events \cite{katsanos2010selection,baker2018improved,jalayer2020intensity}. However, their availability is often constrained, especially for rare events or specific seismic scenarios, and they may not encompass the full spectrum of potential earthquake characteristics. In contrast, synthesized ground motions can be generated from physics-based approaches (e.g., \cite{taborda2015physics,boore2015revisions}) or classical statistics-based approaches (e.g., modulated filtered white noise models \cite{rezaeian2010simulation,su2024review}, spectral-compatible methods \cite{gupta_synthesizing_1993,cacciola_generation_2012,zentner_procedure_2014,yanni2024probabilistic} and many others \cite{mobarakeh_simulation_2002,yamamoto_stochastic_2013}). More recently, deep generative models and physics-informed operator learning frameworks \cite{aquib2024broadband,esfahani2023tfcgan,lehmann2025multiple} have received wide attentions, which complement traditional methods by enabling more flexible and data-driven modeling of complex seismic phenomena. Synthetic motions offer high flexibility, allowing the simulation of a broad array of seismic scenarios with controlled spectral characteristics \cite{kim2024adaptive,su2024importance,mobarakeh_simulation_2002}. Among these synthetic approaches, spectral-compatible stochastic ground motion models (SGMMs) \cite{gupta_synthesizing_1993,cacciola_generation_2012,zentner_procedure_2014,yanni2024probabilistic} are commonly used  when recorded data are insufficient, yet their reliability in predicting structural response remains a key concern \cite{katsanos2010selection,kwong2015selection}. This study evaluates spectral-compatible approaches by comparing recorded and synthetic motions with similar spectral characteristics, including spectral median, variance, and correlation structure.

A comparison between spectral-compatible recorded and synthetic ground motions is essential to evaluate their respective strengths and limitations, particularly in the context of various structural response metrics. A recent study \cite{rezaeian2024findings} offers a comprehensive review of ground motion model validation, classifying validation metrics into two broad categories: ground motion characteristics and structural responses. Ground motion characteristics encompass waveform properties, response spectral intensity measures, peak parameters, and other intensity measures. In contrast, structural response metrics focus on the behavior of both idealized \cite{burks2014validation,galasso2020validation} and realistic structural models \cite{fayaz2021methodology,munjy2022validation}, considering quantities such as maximum inter-story drift ratios, peak floor accelerations, and energy-based measures \cite{chandramohan2016quantifying,tsioulou2019validation,karimzadeh2019use}. Quantitative statistics—including median, dispersion, correlation, and kurtosis—can be evaluated for any scalar validation metric using suites of historical earthquake records alongside comparable synthetic simulations.

While previous studies \cite{galasso2013validation,pei2014variability,karimzadeh2017assessment,li2021evaluation} have explored specific metrics, such as peak response distributions or spectral accelerations, comprehensive evaluations spanning a broader range of stochastic response metrics remain limited. Moreover, as seismic design increasingly integrates uncertainty quantification (UQ), variance-based sensitivity analysis has emerged as a robust tool for identifying key contributors to response variability \cite{sobol2001global,padgett2007sensitivity,lamprou2013life}. This enables engineers to prioritize critical uncertainties while simplifying those with lesser impacts on seismic performance. Despite its potential, limited research has applied variance-based sensitivity analysis to compare the influence of different uncertainty sources on structural responses under synthetic versus recorded excitations \cite{erazo2016uncertainty,kim2025uncertainty,smith2024uncertainty}.

In this context, several prior studies have considered the effects of matched response spectra on structural response but often with limited scope. For example, Bijelic et al. \cite{bijelic2014seismic} evaluated the seismic response of a tall building under recorded and simulated motions, while Iervolino et al. \cite{iervolino2010spectral} assessed the role of spectral shape using artificial and real ground motions for single-degree-of-freedom systems. More recently, Bijelic et al. \cite{bijelic2018validation} examined fragility-based collapse performance for various structures using physics-based synthetic motions, but with a limited number of records and without explicit analysis of spectral correlation. Notably, these studies did not match full spectral correlation matrices across periods. To our knowledge, no prior study has simultaneously matched and evaluated spectral median, variance, and correlation in synthetic and recorded datasets, nor assessed their impact across a comprehensive range of stochastic structural response metrics.

This study aims to fill this gap by systematically comparing stochastic seismic response metrics between synthetic and recorded ground motions that are calibrated to the same unconditional target spectrum. Specifically, five response metrics are investigated: probability distributions, statistical moments and normalized moments (mean, variance, skewness, kurtosis), correlations, tail indices, and variance-based sensitivity indices. Both synthetic and recorded ground motion datasets are calibrated to an equivalent target spectrum, maintaining alignment in spectral median, variance, and correlation characteristics. The sensitivity analysis incorporates both ground motion variability and structural parameter variability, complemented by a parametric study exploring the effects of input parameter distributions on sensitivity patterns. These metrics collectively capture various aspects of seismic response variability, ranging from global characteristics (e.g., distributions and correlations) to extreme behaviors (e.g., tail indices and higher-order moments) and sensitivity patterns (e.g., the relative influence of uncertainties between ground motions and structures). Two archetypal structures are analyzed: a 12-story medium-period building and a 110-story long-period tower. These structures represent distinct dynamic regimes, facilitating an evaluation of how ground motion characteristics influence structural responses across varying periods. By systematically comparing these metrics, the study provides critical insights into the strengths and limitations of synthetic and recorded ground motions in the context of seismic design and risk assessment.


This paper is organized as follows: Section~\ref{Background} introduces the five stochastic seismic response metrics, the ground motion datasets, and the two structural models. Section~\ref{Response_comparison} presents a comparative evaluation of  seismic responses for both ground motion types across the two structures, highlighting key trends and insights. Section~\ref{Summary} synthesizes the results and summarizes the main findings. Finally, Section~\ref{Conclusion} discusses the implications of the findings for seismic design and risk assessment.

\section{Background and comparative framework} \label{Background}

This section outlines the framework for comparing stochastic seismic responses from synthetic and recorded ground motions. Five key metrics are introduced to evaluate global and extreme structural responses, followed by descriptions of the ground motion datasets and structural models employed in the analysis.

\subsection{Input uncertainties in seismic analysis} \label{Uncertainty_sources}

\noindent Seismic responses, denoted as $\vect{Y}=[Y_1,...,Y_{m}]$, typically represent vectors of engineering demand parameters (EDPs), such as inter-story drift ratios (IDRs) and peak floor accelerations (PFAs), which correspond to peak values during seismic events. These responses are influenced by uncertain input variables, including both structural parameters and ground motion excitation, which propagate through nonlinear response history analysis (NLRHA). Mathematically, the seismic response can be expressed as:
\begin{equation}  \label{Eq:Model}
\vect{Y}=\mathcal{M}(\vect{X}_{\mathbf{GM}}(t, \omega), \vect{X}_{\mathbf{S}}) \,,
\end{equation}
where $\mathcal{M}$ denotes the structural response function, $\vect{X}_{\mathbf{S}}$ represents the uncertain structural parameters, and $\vect{X}_{\mathbf{GM}}(t, \omega)$ characterizes the ground motion excitation as a stochastic process indexed by time $t$ and a realization $\omega \in \Omega$, where $\Omega$ is the space of possible ground motion realizations. This formulation explicitly accounts for the inherent randomness of seismic excitations, in which each realization $\omega$ corresponds to a distinct ground motion time history, potentially differing in frequency content, and duration.

Input uncertainties can be categorized into two types: aleatory and epistemic uncertainties. Within the constraints of the models and data used in this analysis, aleatory uncertainties represent the \emph{inherent} variability in ground motions, represented by the stochastic process $\vect{X}_{\mathbf{GM}}(t, \omega)$, arising from the randomness of seismic excitations intrinsic to earthquake phenomena. Epistemic uncertainties arise from incomplete knowledge of structural parameters, denoted as $\vect{X}_{\mathbf{S}}=[X_1,...,X_{n_S}]$. These uncertainties are influenced by variability in material properties, modeling assumptions, and structural configurations.

The complete input uncertainty representation is thus given by $\vect{X} = (\vect{X}_{\mathbf{GM}}(t, \omega), \vect{X}_{\mathbf{S}})$. Modeling these inputs using both synthetic and recorded ground motions enables a direct comparison of their respective impacts on structural responses.

\subsection{Metrics for stochastic seismic responses} \label{Seismic_metrics}

\noindent Five key metrics are employed to systematically compare stochastic seismic responses under synthetic and recorded ground motions. These metrics are evaluated for each EDP $Y_j$, but for simplicity, the subscript $j$ is omitted in the following descriptions:

\begin{enumerate}
    \item \textbf{Probability distributions}: The overall spread and range of EDPs are characterized using probability distributions. To quantify differences between distributions, a quantile-based discrepancy measure, $DF_Q$, is defined as:
    \begin{equation}  \label{Eq:DF_quantile}
    DF_Q = \frac{\frac{1}{N} \sum_{i=1}^{N}\left|Q_{i/N}^{\textrm{synthetic}} - Q_{i/N}^{\textrm{recorded}} \right|}{\mu^{\textrm{synthetic}}} \,,
    \end{equation}
where $Q_{i/N}^{\textrm{synthetic}}$ and $Q_{i/N}^{\textrm{recorded}}$ are the quantiles at rank $i$ in the sorted synthetic and recorded response datasets, $\mu^{\textrm{synthetic}}$ is the mean of the synthetic responses, and $N$ is the dataset size. Smaller $DF_Q$ values indicate greater similarity between the response distributions. Note that $DF_Q$ captures differences in the tails, as quantile deviations are typically more pronounced in tail regions.
    
    \item \textbf{Statistical moments}: Moments and normalized moments, such as mean, variance, skewness, and kurtosis, characterize the central tendency, variability, asymmetry, and tail extremity of EDPs. Differences in these moments are quantified as:    
    \begin{equation}  \label{Eq:DF_moment}
    DF_{M_k} = \frac{\left|M_k^{\textrm{synthetic}} - M_k^{\textrm{recorded}} \right|}{M_k^{\textrm{synthetic}}} \,,
    \end{equation}
    where $M_k$ ($k=1,...,4$) represents the mean, variance, skewness, and kurtosis, respectively.
    
    \item \textbf{Correlations}: Correlations between EDPs quantify the interdependencies among different structural responses, capturing relationships between response modes. For a given pair of EDPs, the difference in correlation estimated from synthetic and recorded ground motions is measured as:
    \begin{equation}  \label{Eq:DF_corr}
    DF_{\rho} = \frac{\left|\rho^{\textrm{synthetic}} - \rho^{\textrm{recorded}} \right|}{\rho^{\textrm{synthetic}}} \,,
    \end{equation}
    where $\rho$ is the Pearson correlation coefficient.

    \item \textbf{Tail indices}: Tail indices characterize the extremity of response distributions, which is critical for evaluating rare-event probabilities. The Hill's estimator \cite{huisman2001tail} is adopted to quantify tail behavior:
    \begin{equation}  \label{Eq:Hill}
    T_{k} = \frac{1}{\frac{1}{k} \sum_{i=1}^{k} \left(\ln{y_{(i)}} - \ln{y_{(k+1)}} \right)} \,,
    \end{equation}
    where $y_{(i)}$ represents the $i$-th largest value in the ordered dataset, and $k$ is the number of extreme data points considered. Smaller $T_k$ values indicate heavier tails. Differences in tail indices are measured as:    
    \begin{equation}  \label{Eq:DF_tail}
    DF_{T_k} = \frac{\left|T_k^{\textrm{synthetic}} - T_k^{\textrm{recorded}} \right|}{T_k^{\textrm{synthetic}}} \,.
    \end{equation}

    \item \textbf{Variance-based sensitivity indices}: Sobol’ sensitivity indices \cite{sobol2001global,saltelli2008global} are adopted to quantify the contributions of ground motion variability and structural parameter uncertainty to response variability. The group-wise Sobol’ index measures the proportion of output variability attributable to a group of input variables $\vect{X}_{\mathbf{u}}$:
    \begin{equation}  \label{Eq:SobolFirst}
    S_\mathbf{u} = \frac{\mathbb{V}\text{ar}_{\vect{X}_{\mathbf{u}}} \left[\mathbb{E}_{\vect{X}_{\sim \mathbf{u}}}(Y|\vect{X}_{\mathbf{u}}) \right]}{\mathbb{V}\text{ar}[Y]} \,,
    \end{equation}
    where $\vect{X}_{\sim \mathbf{u}}$ denotes all input variables except those in group $\mathbf{u}$, and $\mathbb{E}_{\vect{X}_{\sim \mathbf{u}}}$ and $\mathbb{V}\text{ar}_{\vect{X}_{\mathbf{u}}}$ represent the conditional expectation and variance over $\vect{X}_{\sim \mathbf{u}}$ and $\vect{X}_{\mathbf{u}}$, respectively. In this study, $\vect{X}_{\mathbf{u}}$ corresponds to either a group of structural parameters, $\vect{X}_{\mathbf{S}}$, or a set of random variables representing the discretized ground motion stochastic process, given by $\vect{X}_{\mathbf{GM}} = [a(t_1, \omega), a(t_2, \omega), ..., a(t_n, \omega)]$, where $a(t, \omega)$ denotes the ground acceleration at time $t$ for a given realization $\omega$. This formulation enables the decomposition of response variability into aleatory and epistemic contributions. Differences in sensitivity indices are quantified as:   
    \begin{equation}  \label{Eq:DF_sensitivity}
    DF_{S_\mathbf{u}} = \frac{\left|S_\mathbf{u}^{\textrm{synthetic}} - S_\mathbf{u}^{\textrm{recorded}} \right|}{S_\mathbf{u}^{\textrm{synthetic}}} \,.
    \end{equation}
\end{enumerate}
These metrics collectively capture both global response characteristics and extreme behaviors, enabling a insightful comparison between synthetic and recorded ground motions. The comparative results are presented in Section~\ref{Response_comparison}.

\subsection{Synthetic and recorded ground motion datasets} \label{GM_datasets}

\noindent Two ground motion datasets—synthetic and recorded—are developed to facilitate a comparative analysis of seismic responses. The GMM-based unconditional target spectra is adopted in this study. Both datasets are calibrated to a common target spectrum derived from the Ground Motion Prediction Equation (GMPE) by \cite{boore2014nga} and the spectral correlation model by \cite{baker2008correlation}. This target spectrum represents a moderate seismic hazard scenario, serving as a benchmark for ensuring consistency between the datasets. Table~\ref{Tab_TS_v1} summarizes the seismic hazard parameters defining the target spectrum.

The recorded dataset consists of 2,000 ground motions selected from the PEER Next Generation Attenuation (NGA)-West 2 database \cite{ancheta2014nga}. The synthetic dataset comprises 2,000 ground motions generated using a spectral-compatibility method \cite{yanni2024probabilistic}, calibrated to match the statistical characteristics of the target spectrum. Ground motion selection and generation follow standardized procedures \cite{baker2018improved, yanni2024probabilistic}, as detailed in~\ref{App:GM_algo}. Each ground motion is treated as a discretized realization of the ground motion stochastic process, which is used in the comparative analysis.

It is important to note that the synthetic ground motions are generated using the evolutionary spectrum-based formulation of Yanni et al. (2024) \cite{yanni2024probabilistic}, which employs a probabilistic time–frequency modulating envelope and an iterative spectral correction scheme. This variant does not rely on seed recorded accelerograms and is independent of the availability of historical earthquake records. Instead, it ensures consistent enforcement of the target spectral median, variance, and correlation across periods, which are critical to the comparative analysis conducted in this study.

\begin{figure}[H]
  \centering
  \includegraphics[scale=0.48] {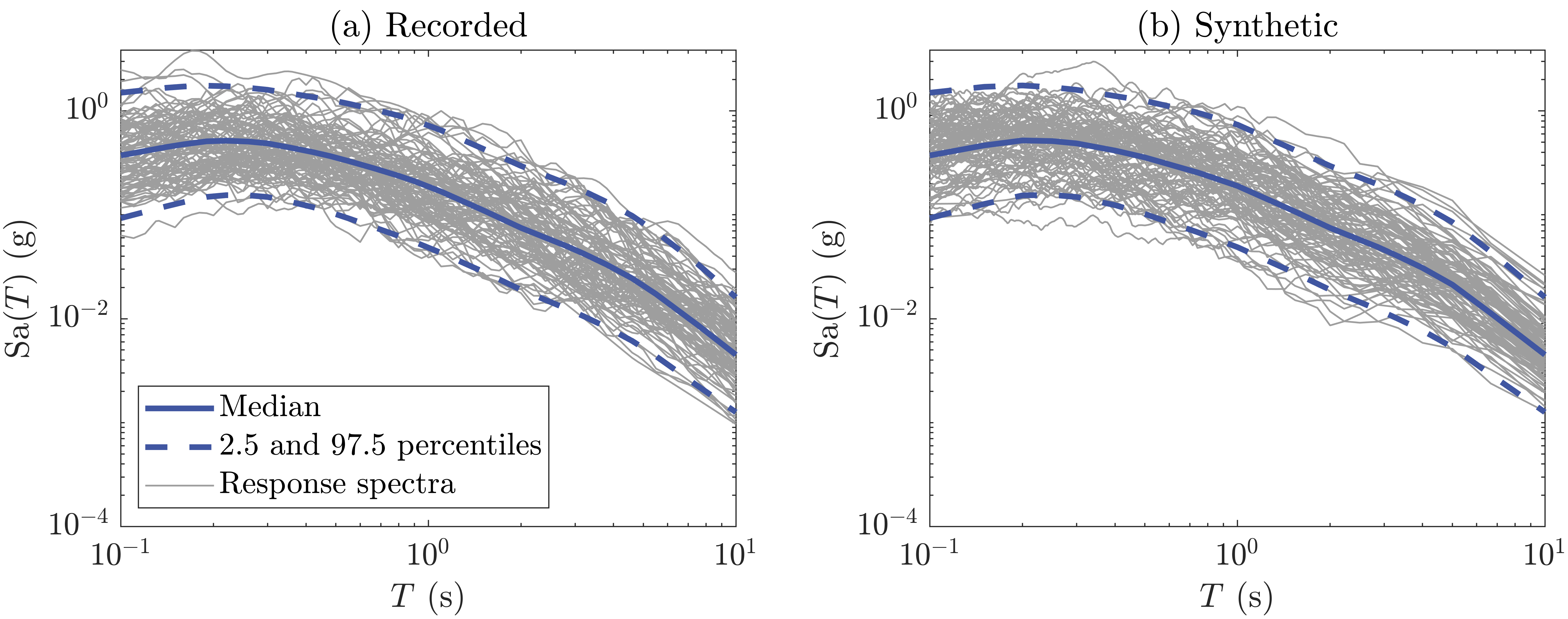}
  \caption{\textbf{Response spectra of (a) recorded and (b) synthetic ground motions matching the target spectrum}. Each plot shows 100 response spectra, with the median and 2.5\%-97.5\% quantiles of the target spectrum superimposed.}
  \label{Fig_GM_spectra}
\end{figure}

\begin{figure}[H]
  \centering
  \includegraphics[scale=0.48] {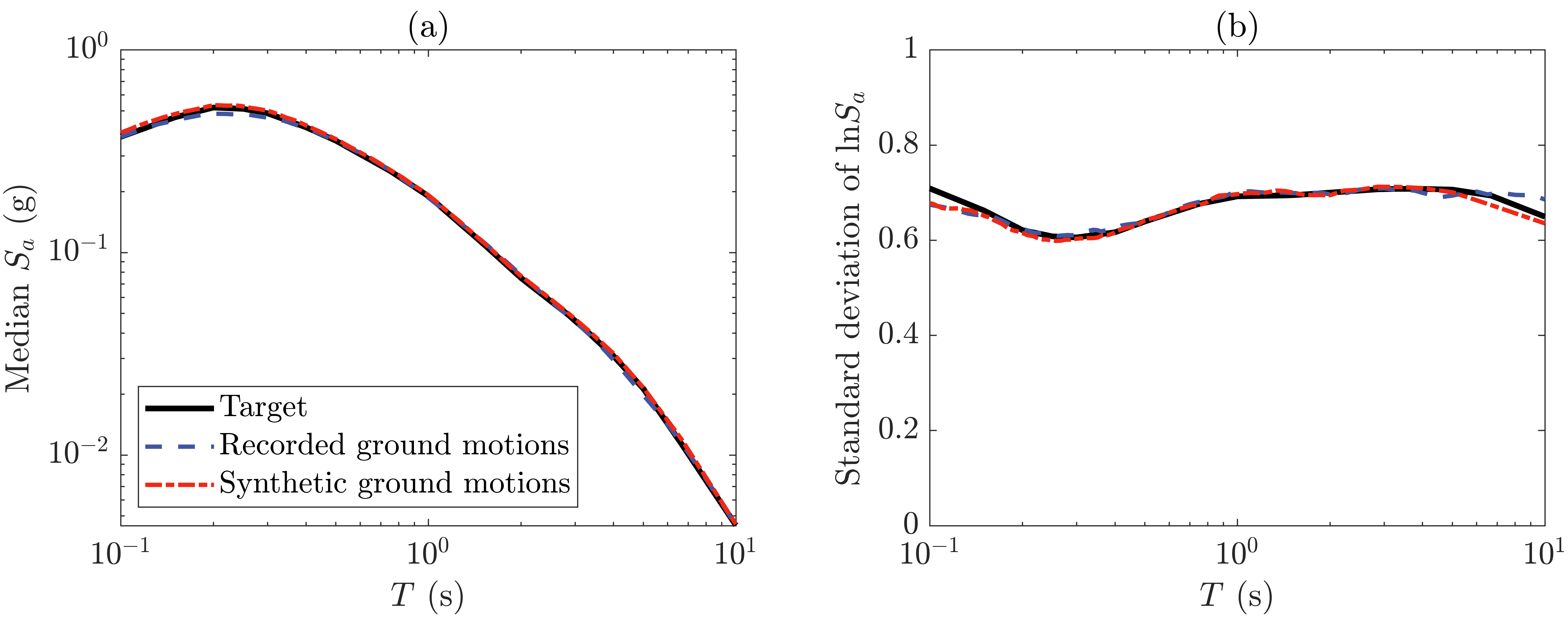}
  \caption{\textbf{Comparison of the (a) spectral median and (b) variability between the target spectrum and the recorded/synthetic ground motions}. Statistics are computed using 2,000 ground motions in each dataset.}
  \label{Fig_TS_compre}
\end{figure}

\begin{figure}[H]
  \centering
  \includegraphics[scale=0.46] {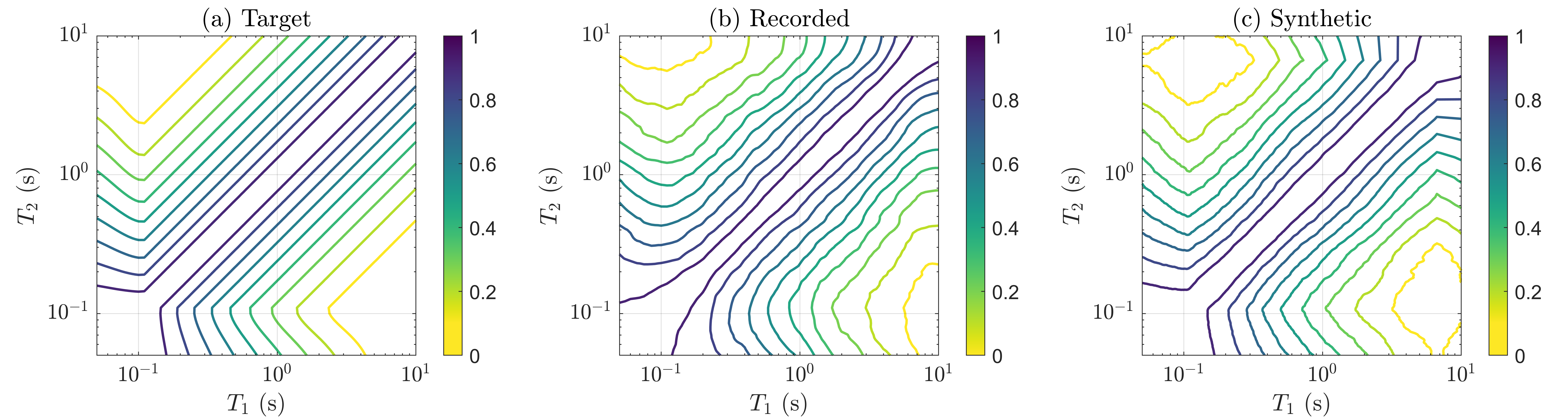}
  \caption{\textbf{Spectral correlations: (a) target correlation structure, (b) empirical correlations from recorded motions, and (c) empirical correlations from synthetic motions}. Contours represent correlation coefficients between spectral accelerations at periods $T_1$ and $T_2$. Panels (b) and (c) are computed using 2,000 ground motions in each dataset.}
  \label{Fig_TS_compre_corr}
\end{figure}

Figure~\ref{Fig_GM_spectra} presents the response spectra reconstructed from the recorded and synthetic ground motions. Figure~\ref{Fig_TS_compre} and Figure~\ref{Fig_TS_compre_corr} compare their spectral median, variability, and correlations against the target spectrum, confirming the equivalence of spectral characteristics between the two datasets. This calibration ensures that any observed differences in seismic response metrics reflect intrinsic disparities in ground motion characteristics rather than spectral inconsistencies, thereby establishing a robust basis for comparison.

\begin{table}[H]
  \caption{\textbf{Seismic hazard parameters for the target spectrum}.}
  \label{Tab_TS_v1}
  \centering
  \begin{tabular}{l l}
    \toprule
    Parameter & Value \\
    \midrule
    Earthquake magnitude & 6.5 \\
    Closest distance to fault rupture (km) & 10 \\
    Average shear wave velocity in the top 30 m (m/s) & 450 \\
    Fault type & Normal \\
    Region & California \\
    \bottomrule
  \end{tabular}
\end{table}

\begin{figure}[H]
  \centering
  \includegraphics[scale=0.52] {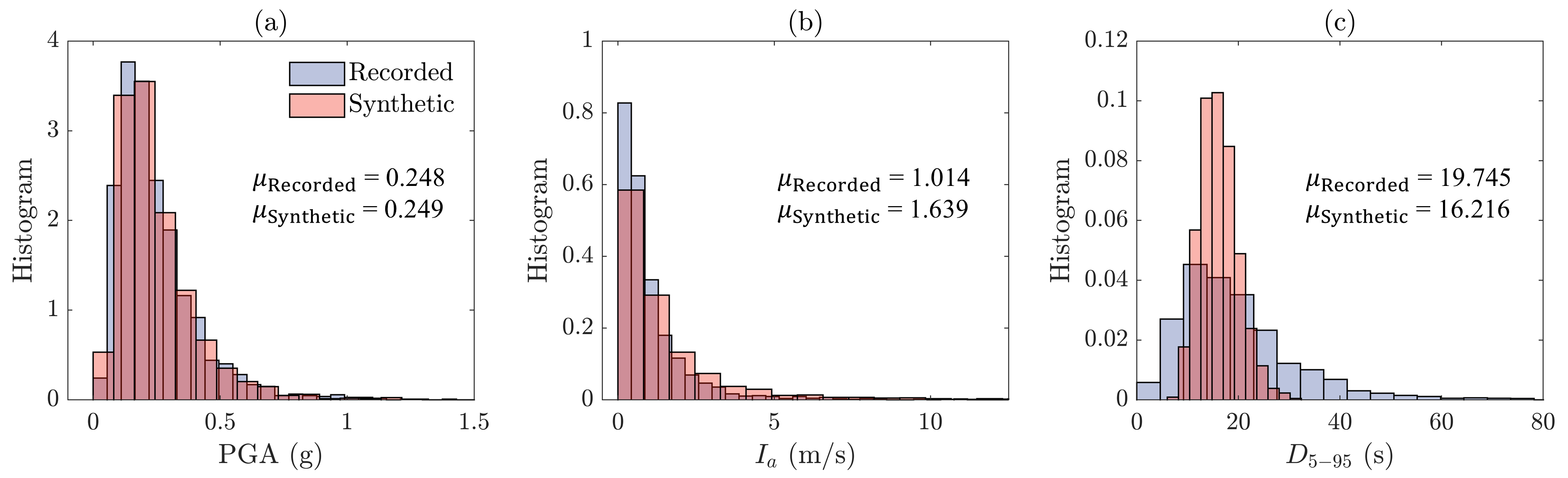}
  \caption{\textbf{Comparison of ground motion intensity measures between datasets: (a) PGA, (b) Arias intensity, and (c) significant duration.} Histograms are based on 2,000 ground motions for each dataset.}
  \label{Fig_IM_hist}
\end{figure}

In addition, Figure~\ref{Fig_IM_hist} summarizes key ground motion intensity measures across the recorded and synthetic datasets, including peak ground acceleration (PGA), Arias intensity ($I_a$), and significant duration ($D_{5–95}$, defined as the time interval between 5\% and 95\% of Arias intensity accumulation). The mean values of each intensity measure are also reported. Notably, PGA and $I_a$ exhibit comparable distributions across the datasets, while $D_{5–95}$ is shorter for the synthetic motions, with smaller variability. It is also noted that within the recorded dataset, 418 motions are identified as near-fault events (defined by rupture distances less than 15km), and 118 motions exhibit characteristics commonly classified as pulse-like, according to the PEER database. The complete datasets are available through an open repository, as described in the Data Availability section.

\subsection{Structural models} \label{Structural_models}

\noindent Two structural models are employed to investigate the influence of ground motion characteristics on seismic responses across distinct dynamic regimes: a 12-story medium-period building and a 110-story long-period tower. These models are selected to represent typical mid-rise and high-rise multi-degree-of-freedom (MDOF) structures, enabling a comparative evaluation of global and extreme response characteristics under synthetic and recorded ground motions.

\begin{itemize}
    \item \textbf{Medium-period building}: A 12-story shear-type MDOF system with a fundamental period of 1.09 seconds. The seismic response of this archetype is predominantly governed by low-mode dynamics and reflects typical behavior of mid-rise structural systems.

    \item \textbf{Long-period tower}: A limiting case of 110-story shear-type MDOF system with a fundamental period of 9.69 seconds. Its seismic responses are significantly influenced by higher modes and low-frequency ground motion components\footnote{Although this case is presented as a limiting example of a very long-period structure, its structural system does not correspond to that of a realistic structure of this height.}.

\end{itemize}
These models are not intended to replicate specific real buildings but rather to represent distinct dynamic regimes — one dominated by fundamental mode response and the other by higher-mode effects and long-period sensitivity. Both structures are idealized as shear-building models with uniformly distributed mass and constant story stiffness. All story heights are set to 3.65 meters. The nonlinear response simulation is conducted using a bilinear hysteretic material model (Steel01 in OpenSees), which includes yielding, unloading–reloading hysteresis, and post-yield hardening. Rayleigh damping is applied with 2\% critical damping in the first two modes. Post-yield stiffness is defined as 5\% of the elastic stiffness. This modeling captures nonlinear deformation, residual drift, and cumulative damage effects arising from duration differences between synthetic and recorded motions. However, the structural models do not account for element-level redistribution, $P$–$\Delta$ effects, or vertical irregularities (e.g., outriggers), allowing for controlled comparisons while acknowledging certain limitations in realism.

The seismic responses of interest include peak interstory drift ratios (IDRs) and peak floor accelerations (PFAs), which are widely used in seismic performance evaluation. Nonlinear time-history analyses are conducted using OpenSees \cite{mckenna2011opensees}.

\begin{figure}[H]
  \centering
  \includegraphics[scale=0.41] {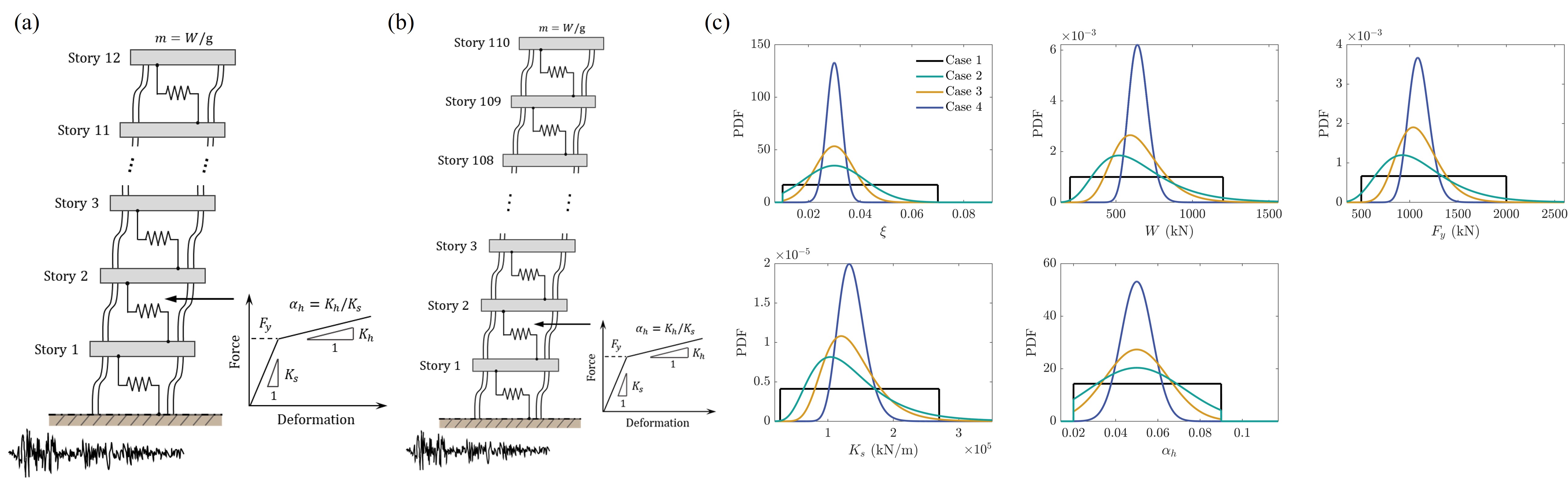}
  \caption{\textbf{Structural models: (a) medium-period building, (b) long-period tower, and (c) PDFs of uncertain structural parameters across four cases}.}
  \label{Fig_MDOF}
\end{figure}

Uncertainties in structural parameters $\vect{X}_{\mathbf{S}}$, including damping ratio, floor weight, yield force, story stiffness, and strength hardening ratio, are modeled using standard distribution models. Table~\ref{Tab_MDOF_rvs} summarizes these distributions across four cases: Case 1 assumes uniform distributions, while Cases 2–4 employ lognormal and truncated Gaussian distributions with decreasing coefficients of variation (c.o.v). These uncertainties are consistently applied to both structural models in the comparative analysis. Figure~\ref{Fig_MDOF} illustrates the structural models and the probability density functions (PDFs) of the parameters. Note that four test cases with different distribution models are considered to ensure the robustness of the comparative study conclusions.

\begin{table}[H]
  \caption{\textbf{Probabilistic distributions of uncertain structural parameters for the structural models}.}
  \label{Tab_MDOF_rvs}
  \centering
  \begin{tabular}{l l l l l}
    \toprule
    Modeling parameter & Case 1 & Case 2 & Case 3 & Case 4 \\
    \cmidrule{1-5}
    Distribution & Uniform & \multicolumn{3}{l}{Lognormal \& Truncated Gaussian} \\
    \midrule
    Damping ratio, $\xi$ & [0.01, 0.07] & $\mu$=0.03 & $\mu$=0.03 & $\mu$=0.03 \\
    & & c.o.v=0.40 & c.o.v=0.25 & c.o.v=0.10 \\
    Floow weight, $W$ (kN) & [200, 1200] & $\mu$=650 & $\mu$=650 & $\mu$=650 \\
    & & c.o.v=0.40 & c.o.v=0.25 & c.o.v=0.10 \\
    Yield force, $F_y$ (kN) & [500, 2000] & $\mu$=1100 & $\mu$=1100 & $\mu$=1100 \\
    & & c.o.v=0.35 & c.o.v=0.20 & c.o.v=0.10 \\
    Story stiffness, $K_s$ (kN/m) & [27000, 270000] & $\mu$=137000 & $\mu$=137000 & $\mu$=137000 \\
    & & c.o.v=0.45 & c.o.v=0.30 & c.o.v=0.15 \\
    Hardening ratio, $\alpha_h$ & [0.02, 0.09] & $\mu$=0.05 & $\mu$=0.05 & $\mu$=0.05 \\
    & & c.o.v=0.45 & c.o.v=0.30 & c.o.v=0.15 \\
    \bottomrule
  \end{tabular}  \\
  \raggedright{\hspace{0.0cm} {\footnotesize * Lower and upper bounds are provided for Case 1, while mean ($\mu$) and c.o.v values are specified for Cases 2-4. For Cases 2-4, truncated Gaussian distributions are applied to $\xi$ and $\alpha_h$, while other parameters follow lognormal distributions.}} \\
\end{table}

\section{Comparative analysis of seismic responses: synthetic versus recorded motions} \label{Response_comparison}

This section presents a detailed comparative evaluation of seismic response metrics for synthetic and recorded ground motions. Five key metrics are analyzed: distributions, statistical moments, correlations, tail indices, and sensitivity indices. The first four metrics focus exclusively on aleatory uncertainties by considering ground motion variability, while structural parameters are fixed at their mean values, as specified in Case 1 of Table~\ref{Tab_MDOF_rvs}, ensuring that the analysis isolates the effects of ground motion variability on seismic responses. The fifth metric incorporates both ground motion variability and structural parameter uncertainty to evaluate their combined effect and relative contributions to seismic response variability. The analysis is conducted for both structural models, examining trends in global and extreme behaviors.

\subsection{Metric 1: Probability distributions} \label{1distribution}

\begin{figure}[H]
  \centering
  \includegraphics[scale=0.395] {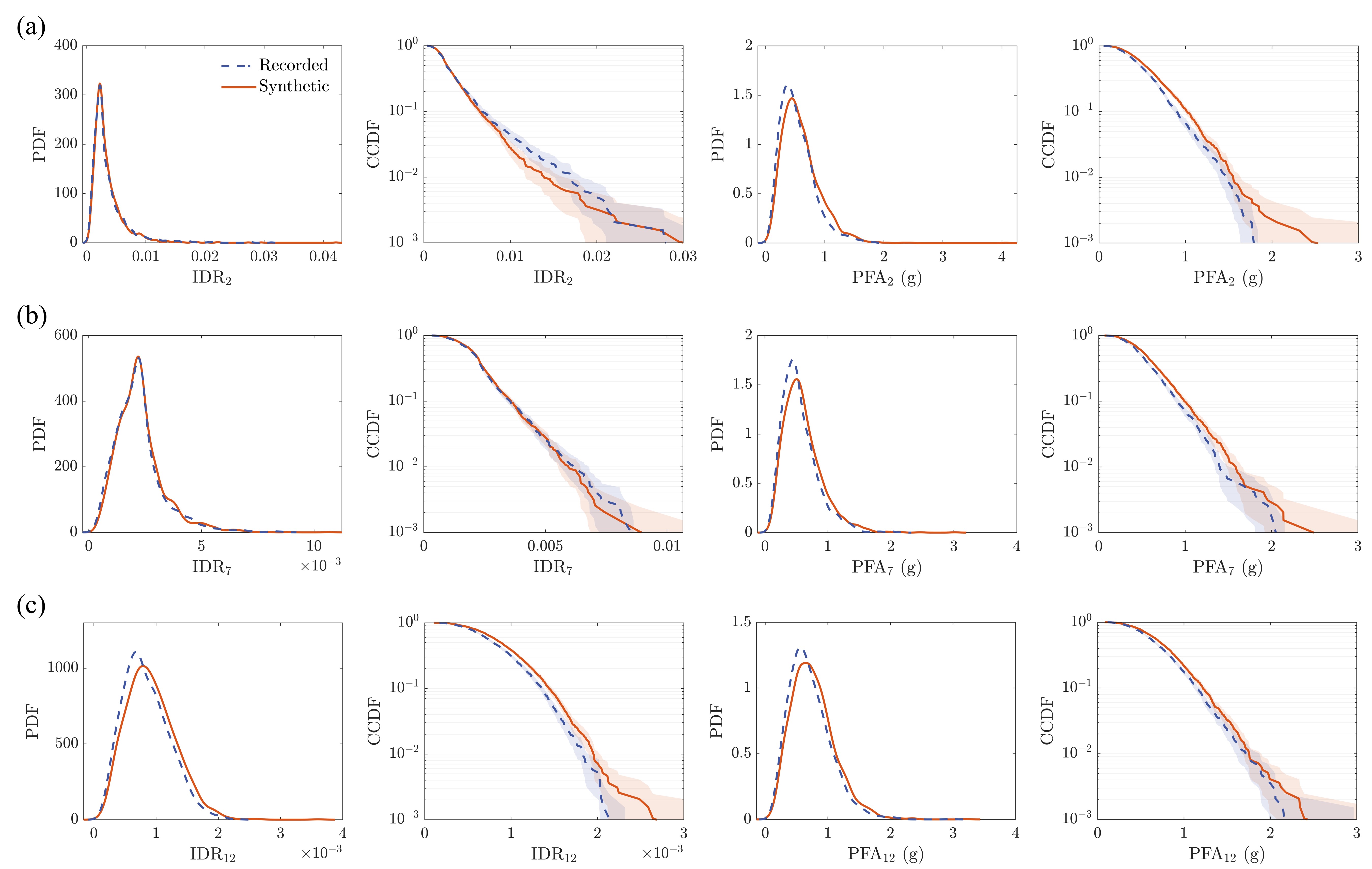}
  \caption{\textbf{Empirical PDFs and CCDFs of seismic responses for the medium-period building: (a) 2nd story, (b) 7th story, and (c) 12th story.} Distribution functions are shown for IDRs and PFAs. For CCDFs, 95\% confidence intervals are depicted as shaded regions.}
  \label{Fig_MDOF12_PDF}
\end{figure}

\begin{figure}[H]
  \centering
  \includegraphics[scale=0.395] {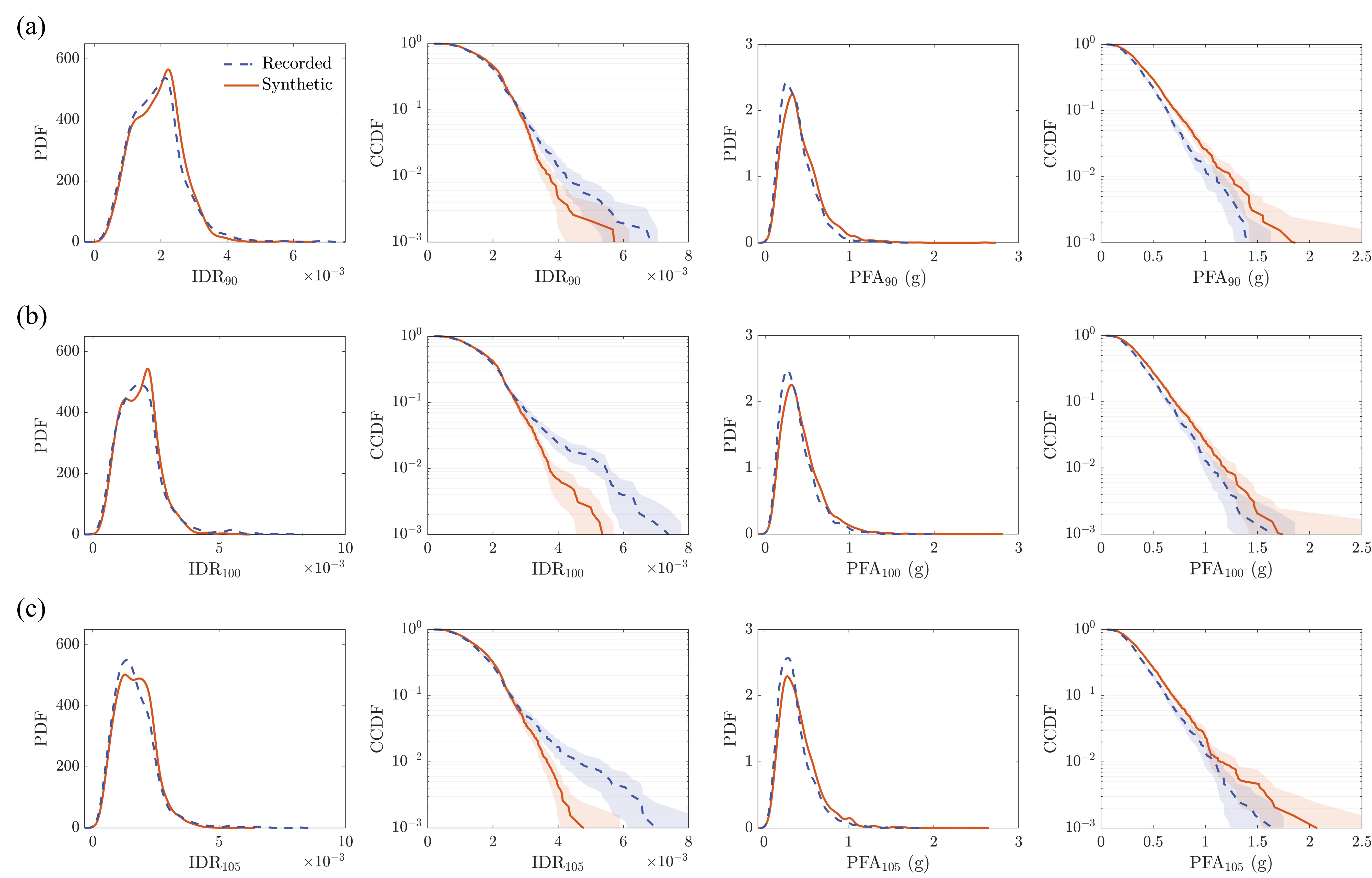}
  \caption{\textbf{Empirical PDFs and CCDFs of seismic responses for the long-period tower: (a) 90th story, (b) 100th story, and (c) 105th story.} Distribution functions are shown for IDRs and PFAs. For CCDFs, 95\% confidence intervals are depicted as shaded regions.}
  \label{Fig_MDOFtall_PDF}
\end{figure}

\noindent Figures~\ref{Fig_MDOF12_PDF} and~\ref{Fig_MDOFtall_PDF} present the empirical PDFs and complementary cumulative distribution functions (CCDFs) of EDPs—peak IDRs and PFAs—for the medium-period building and the long-period tower. Both structural models exhibit close agreement between responses to synthetic and recorded ground motions in the central ranges of the PDFs, indicating comparable global response characteristics under typical seismic scenarios. This alignment suggests that calibrating the synthetic and recorded ground motions to a target spectrum ensures consistency in global response characteristics. However, discrepancies become apparent in the tails of the distributions, particularly for the IDRs of the long-period tower. These differences can be attributed to the non-bijective mapping between spectral acceleration and ground motions, where both selection and generation methods introduce variability in time-domain characteristics despite comparable response spectra.

\begin{figure}[H]
  \centering
  \includegraphics[scale=0.46] {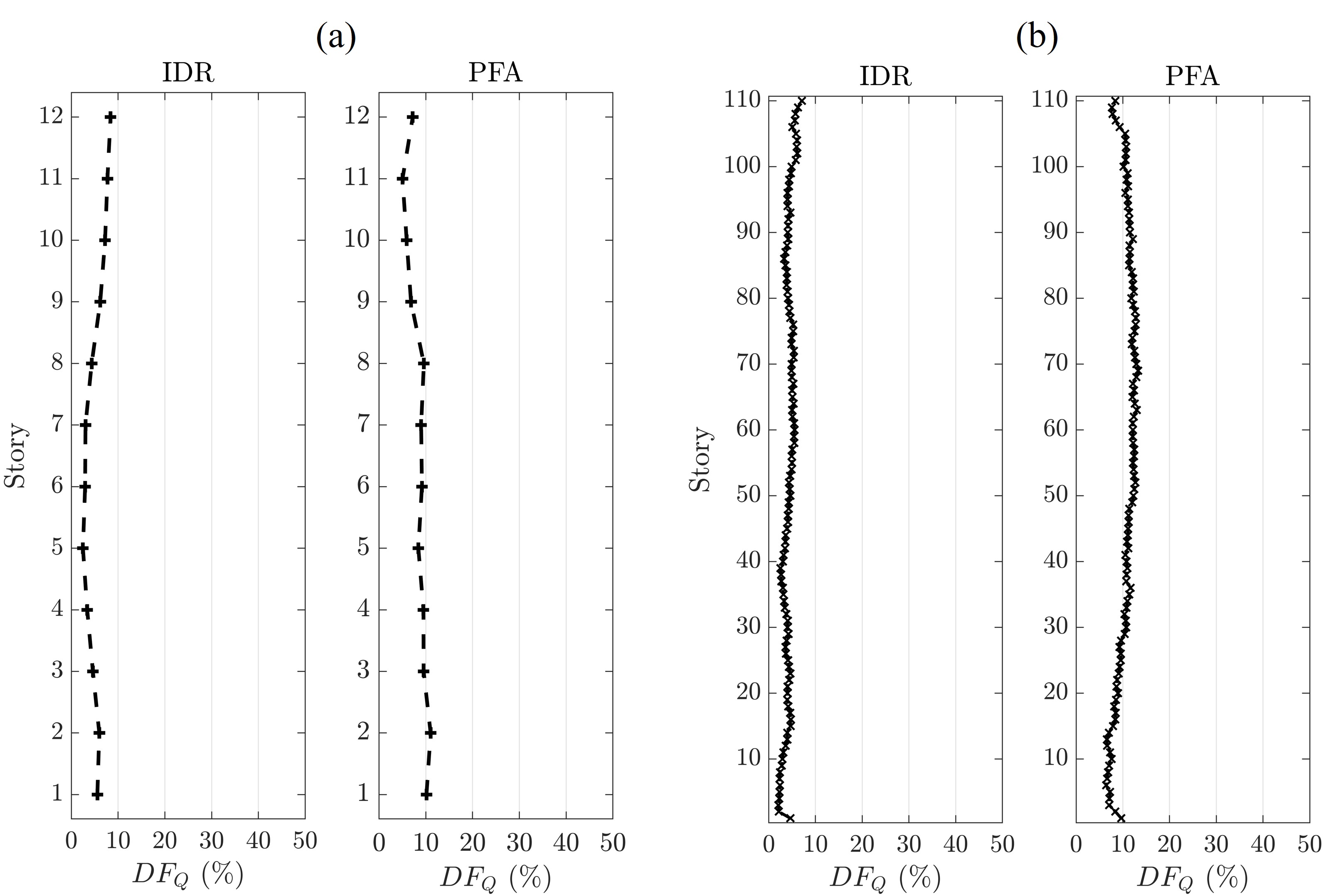}
  \caption{\textbf{Quantile-based distribution differences between recorded and synthetic motion response datasets for (a) the medium-period building and (b) the long-period tower}.}
  \label{Fig_MDOF_DF_Q}
\end{figure}

Figure~\ref{Fig_MDOF_DF_Q} quantifies these discrepancies using the quantile-based distribution difference measure ($DF_Q$). For the medium-period building, $DF_Q$ values range from 2\% to 10\% across all stories, indicating minor differences between the datasets. For the long-period tower, the discrepancies are slightly larger, though they generally remain within 13\%. These findings suggest that synthetic motions effectively capture global distributions but are less reliable in reproducing tail behavior, particularly for structures sensitive to low-frequency ground motion components.


\subsection{Metric 2: Statistical moments}  \label{2moments}

\begin{figure}[H]
  \centering
  \includegraphics[scale=0.48] {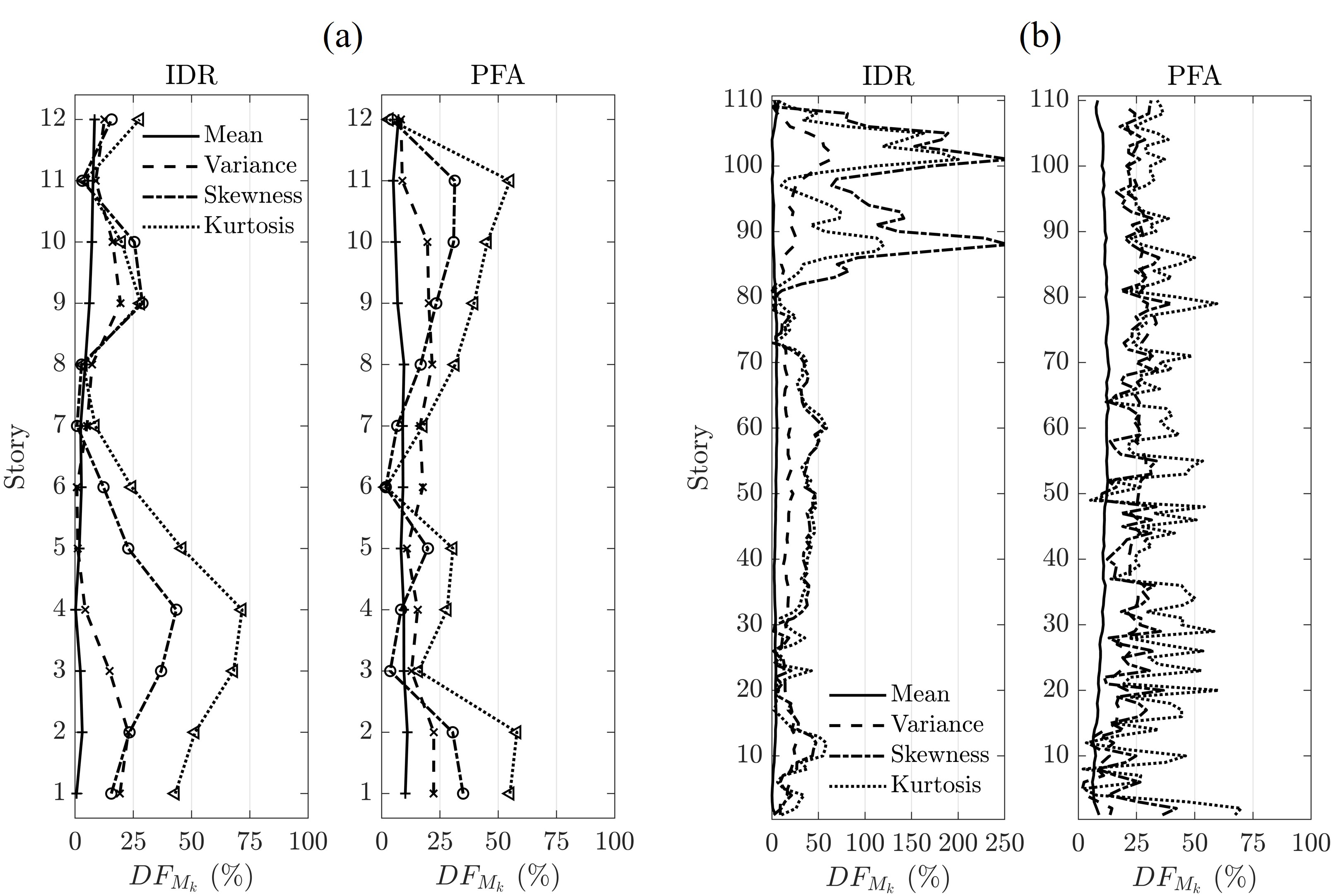}
  \caption{\textbf{Differences in statistical moments (mean, variance, skewness, and kurtosis) between synthetic and recorded response datasets for (a) the medium-period building and (b) the long-period tower.} Results are presented for IDRs and PFAs.}
  \label{Fig_MDOF_DF_M}
\end{figure}

\noindent Figure~\ref{Fig_MDOF_DF_M} illustrates differences in normalized moments ($DF_{M_k}$, $i=k,...,4$)—mean, variance, skewness, and kurtosis—between responses to synthetic and recorded ground motions for both structural models. Differences in mean and variance remain within 3\%–20\%, indicating close agreement in central tendencies and overall variability.  These results demonstrate that well-calibrated synthetic motions can effectively capture general response characteristics.

However, significant differences are observed for higher-order normalized moments, such as skewness and kurtosis, particularly for the long-period tower. For instance, the kurtosis of the IDR distribution at the 100th story of the tower exhibits a discrepancy exceeding 100\%. These findings reveal the limitations of synthetic motions in reproducing extreme variations and outliers present in recorded motions. Synthetic motions, constrained by the Gaussian or Gaussian-like assumptions inherent to SGMMs, tend to smooth variability, while recorded motions often exhibit complex, non-Gaussian features, such as near-fault pulse effects. Additionally, limitations in ground motion selection and generation—such as imperfect spectral matching and variability in time-domain realizations—may accumulate, further amplifying discrepancies in higher-order moments. The heightened sensitivity of long-period structure to low-frequency ground motion components exacerbates these discrepancies, particularly for IDRs, which dominate the dynamic responses of such structures. These results emphasize the need for caution when using synthetic motions to model extreme scenarios.

\subsection{Metric 3: Correlations between EDPs} \label{3correlation}

\begin{figure}[H]
  \centering
  \includegraphics[scale=0.41] {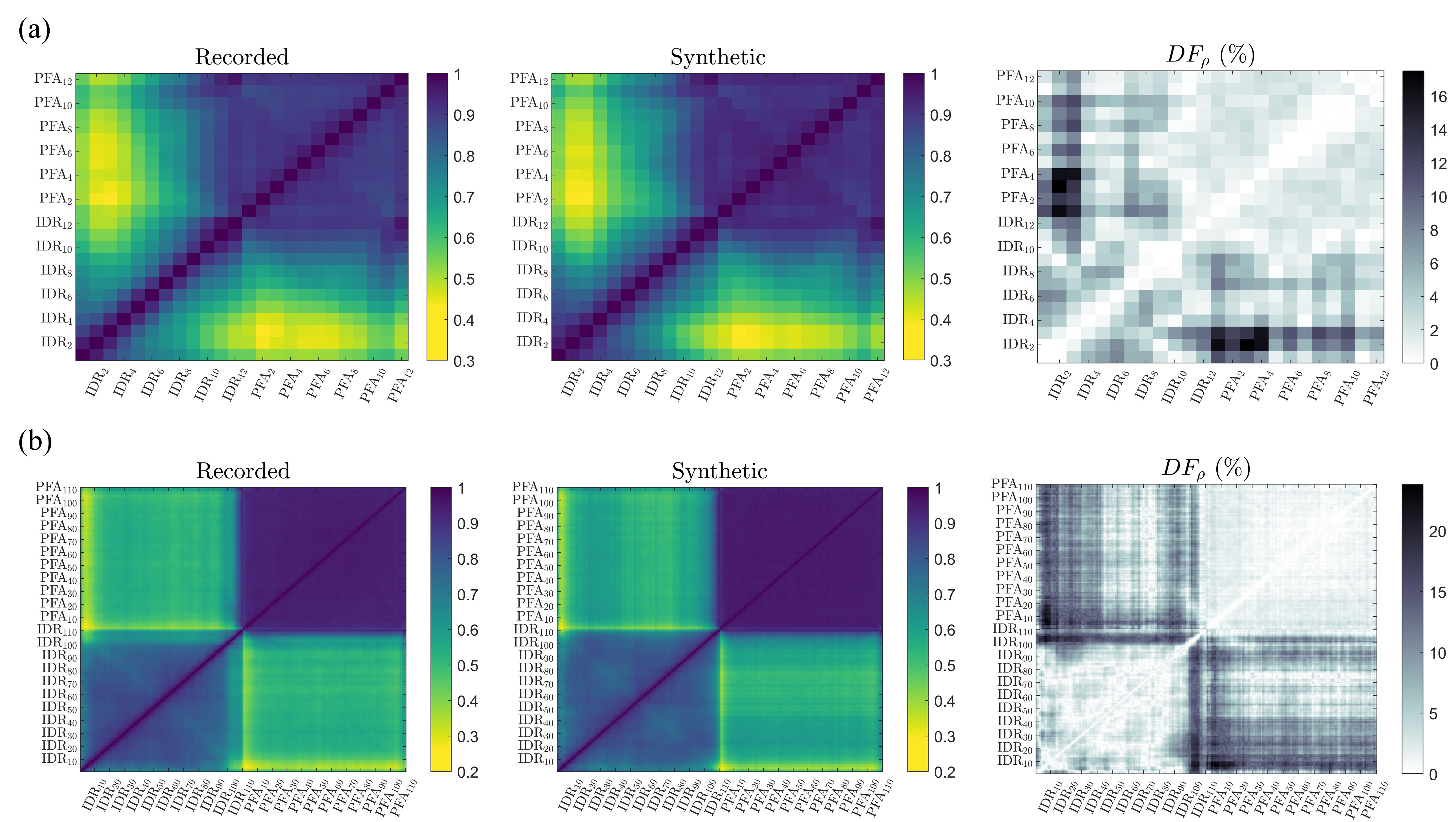}
  \caption{\textbf{Correlation matrices for recorded and synthetic response datasets and their differences for (a) the medium-period building and (b) the long-period tower}.}
  \label{Fig_MDOF_DF_corr}
\end{figure}

\noindent Figure~\ref{Fig_MDOF_DF_corr} presents correlation matrices for EDPs under synthetic and recorded motions, along with the corresponding differences quantified by $DF_{\rho}$. Correlations are computed for all pairs of IDRs and PFAs across both structural models.

For the medium-period building, correlation coefficients show strong consistency between synthetic and recorded datasets, with differences generally below 15\%. Notable discrepancies, such as those observed between IDR$_2$ and PFA$_4$, occur in low-correlation regions, where differences are amplified due to small baseline (denominator) values. These results confirm that synthetic motions effectively replicate the correlation patterns observed in recorded datasets, demonstrating their reliability in capturing global interdependencies. For the long-period tower, correlation differences are slightly larger, reaching up to 20\%. These discrepancies are attributed to complex features in recorded motions, such as pulse-like effects,  which are less pronounced in synthetic datasets. Despite these differences, global correlation patterns are preserved, underscoring the robustness of synthetic motions in capturing interdependencies across EDPs.

\subsection{Metric 4: Tail indices} \label{4tail}

\begin{figure}[H]
  \centering
  \includegraphics[scale=0.48] {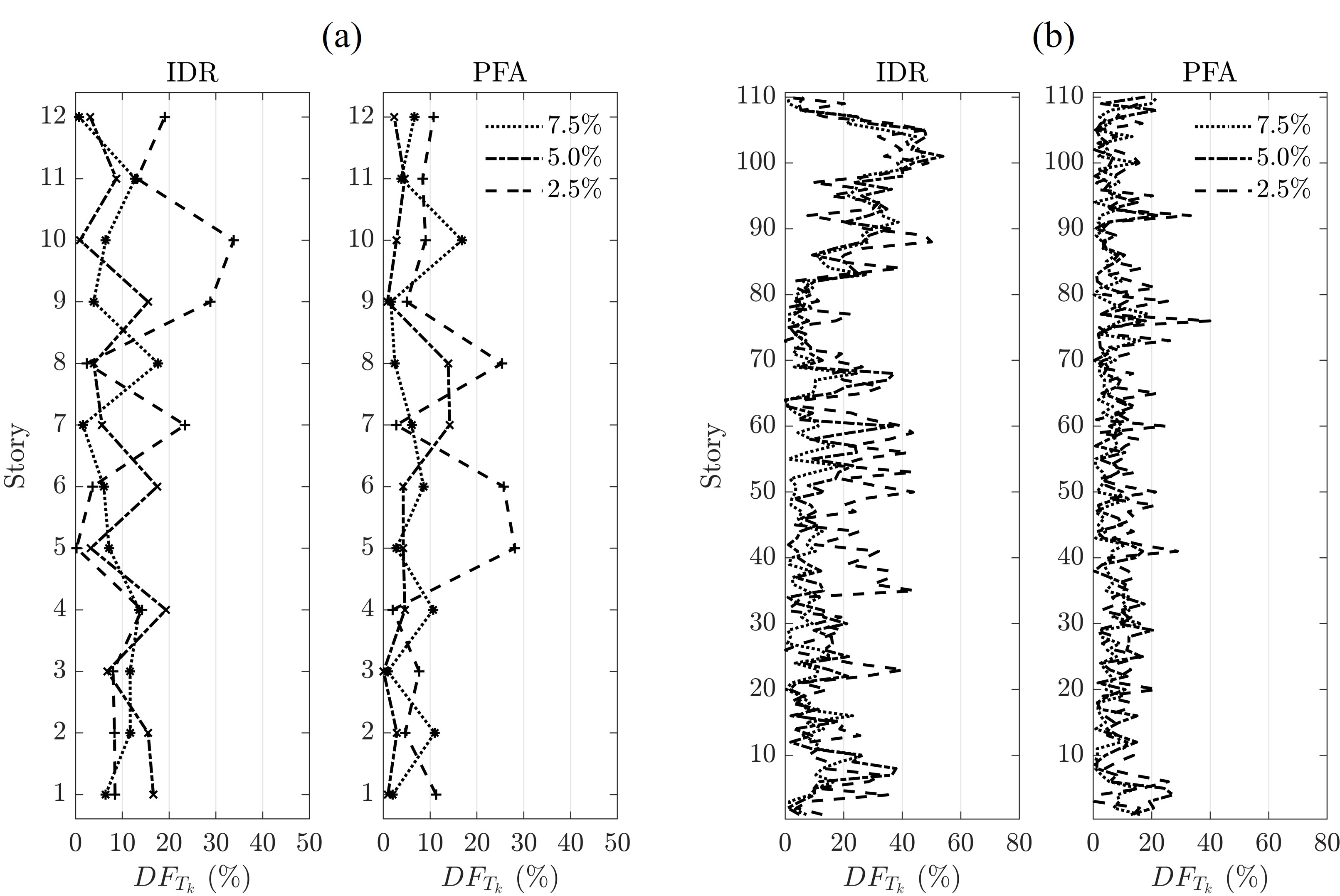}
  \caption{\textbf{Tail index differences between recorded and synthetic motion response datasets for (a) the medium-period building and (b) the long-period tower}.}
  \label{Fig_MDOF_DF_tail}
\end{figure}

\noindent Figure~\ref{Fig_MDOF_DF_tail} compares differences in tail indices ($DF_{T_k}$) for IDRs and PFAs at three exceedance levels: 7.5\%, 5\%, and 2.5\% tails (corresponding to $k=150$, $k=100$, and $k=50$, respectively). As the analysis moves  toward more extreme tails, discrepancies in tail indices become increasingly pronounced. For the medium-period building, differences exceed 30\% at the 2.5\% tail level, while for the long-period tower, discrepancies range from 20\% to 60\%. These results underscore the limitations of synthetic motions in capturing the extreme variability observed in recorded ground motions.

As noted in the higher-order moment differences discussed in Section~\ref{2moments}, the Gaussian-like assumptions inherent in the SGMMs lead to smoothed variability. This smoothing effect contributes to significant differences in the tail behavior of seismic responses. Discrepancies in tail indices also arise from limitations in ground motion selection and generation, such as imperfect spectral matching. Although synthetic motions are calibrated to match spectral characteristics, their inability to capture intricate non-Gaussian features results in attenuated tail behavior. Consequently, synthetic motions may not fully represent rare-event scenarios that are critical for assessing extreme demands on long-period structures.

\subsection{Metric 5: Sensitivity indices to aleatory and epistemic uncertainties} \label{5sensitivity}

\noindent Variance-based sensitivity indices, computed using Saltelli’s method \cite{saltelli2010variance} (details in \ref{App:Saltelli}), quantify the contributions of aleatory (ground motion) and epistemic (structural parameter) uncertainties to seismic responses. This group-wise sensitivity analysis is conducted across the four cases defined in Table~\ref{Tab_MDOF_rvs}, examining how variations in structural parameter distributions affect sensitivity patterns.

\begin{figure}[H]
  \centering
  \includegraphics[scale=0.39] {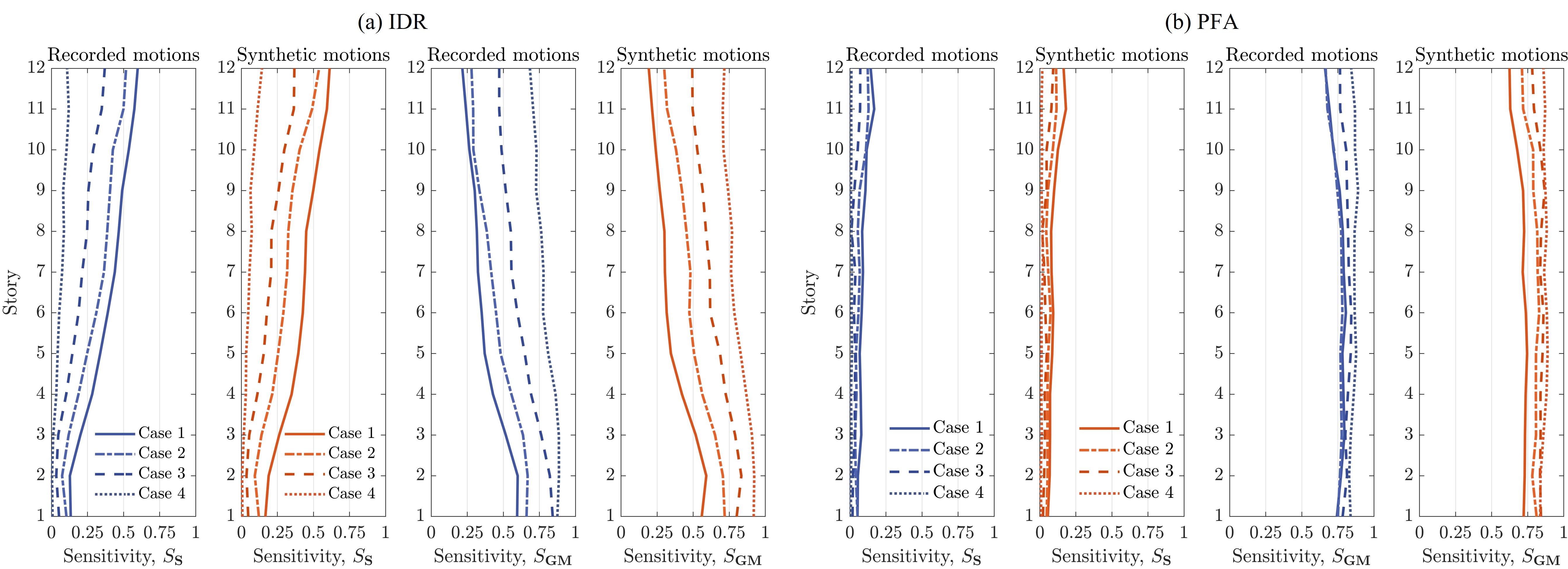}
  \caption{\textbf{Sobol' sensitivity indices for the medium-period building under two groups of input uncertainties: (a) IDRs, and (b) PFAs}. Indices are presented for both structural and ground motion uncertainties across four cases with varying structural variability.}
  \label{Fig_MDOF12_S}
\end{figure}

Tables~\ref{Tab_MDOF_Su_IDR} and~\ref{Tab_MDOF_Su_PFA} present the sensitivity indices for ground motion variability ($S_{\mathbf{GM}}$) and structural parameter variability ($S_{\mathbf{S}}$) for IDRs and PFAs, respectively, for the medium-period building, as illustrated in Figure~\ref{Fig_MDOF12_S}. Sensitivity indices are computed from ten independent runs of the sensitivity analysis algorithm, each using different realizations of the uncertain structural parameters, with the mean values reported in the tables. A clear trend emerges: as the c.o.v of structural parameters decreases (from Case 1 to Case 4), the influence of structural uncertainties diminishes, while ground motion variability becomes increasingly dominant. As building height increases, the relative contribution of structural uncertainties grows, particularly for IDRs in the upper stories. This effect is most pronounced in Case 1, where structural variability is highest, but the pattern holds across all cases. In contrast, for PFAs, ground motion variability remains the dominant factor in all cases, even as structural variability increases. This sensitivity pattern is consistent across both synthetic and recorded datasets.

\begin{figure}[H]
  \centering
  \includegraphics[scale=0.49] {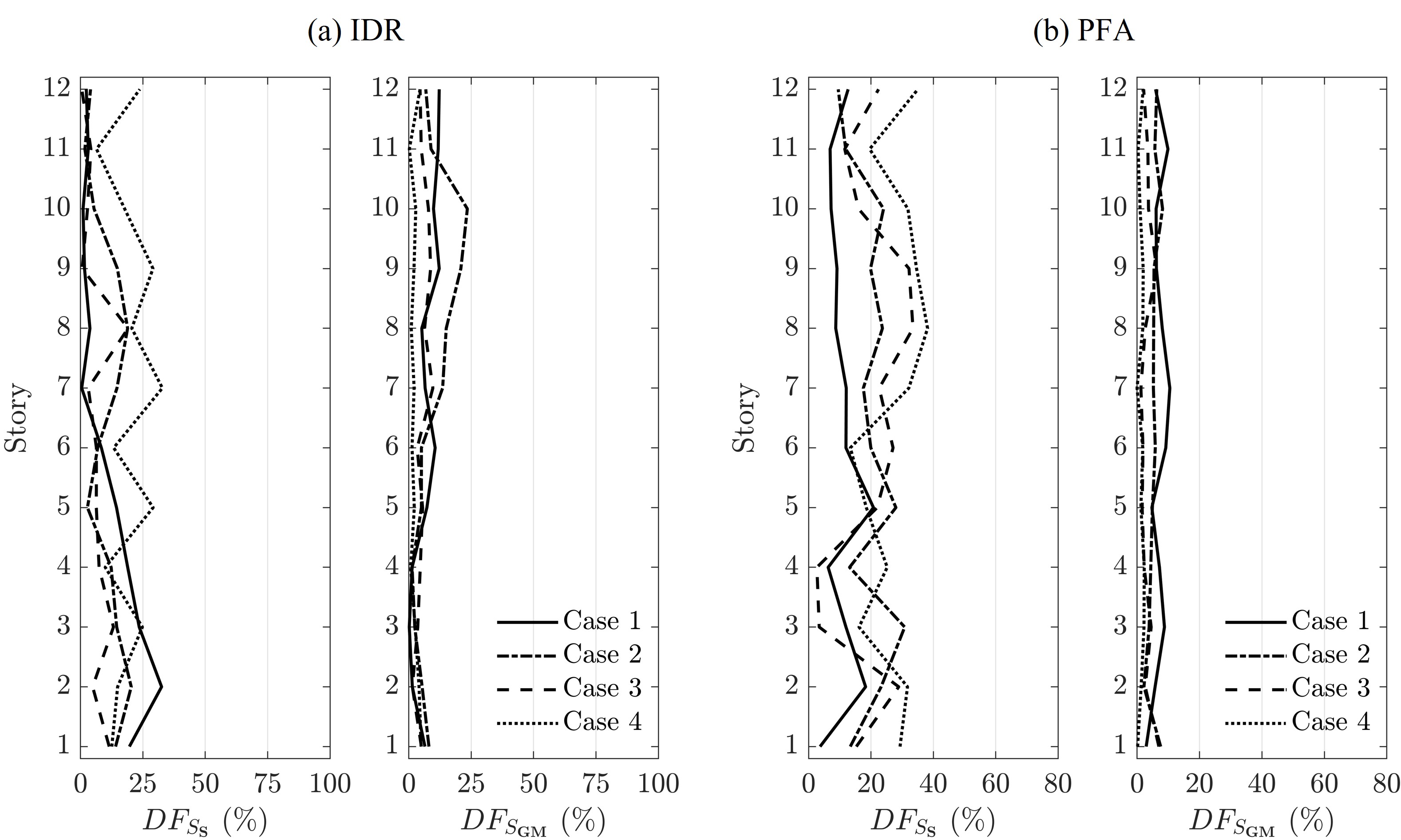}
  \caption{\textbf{Differences in Sobol' sensitivity indices between synthetic and recorded datasets for the medium-period building: (a) IDRs, and (b) PFAs}. Differences are evaluated under the four structural parameter distribution cases.}
  \label{Fig_MDOF_DF_S}
\end{figure}

The corresponding differences in sensitivity indices between synthetic and recorded datasets are shown in Figure~\ref{Fig_MDOF_DF_S}. Overall, these differences remain below 25\%, consistent with the observation that variance-based sensitivity indices reflect global response characteristics. Notably, the relative rankings and variation patterns of the sensitivity indices align closely between synthetic and recorded datasets, demonstrating the robustness of synthetic motions in capturing global sensitivity trends.

\begin{figure}[H]
  \centering
  \includegraphics[scale=0.39] {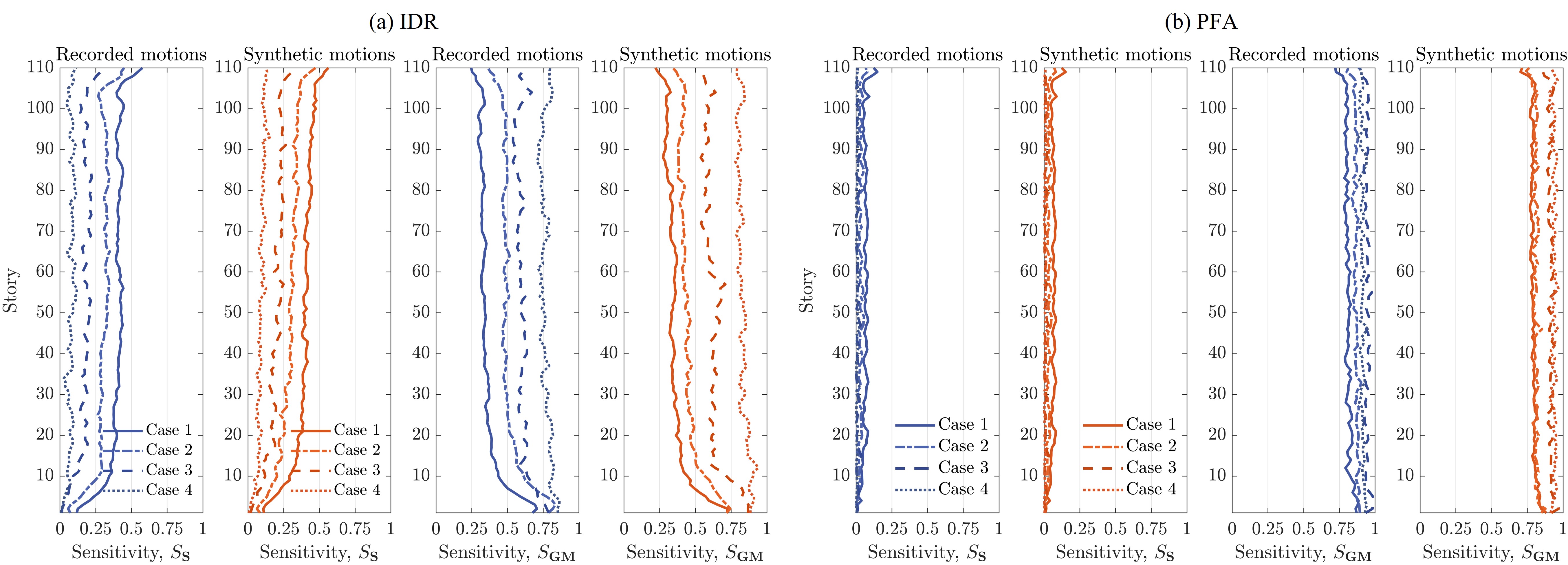}
  \caption{\textbf{Sobol' sensitivity indices for the long-period tower under two groups of input uncertainties: (a) IDRs, and (b) PFAs}. Indices are presented for both structural and ground motion uncertainties across four cases with varying structural variability.}
  \label{Fig_MDOFtall_S}
\end{figure}

\begin{figure}[H]
  \centering
  \includegraphics[scale=0.49] {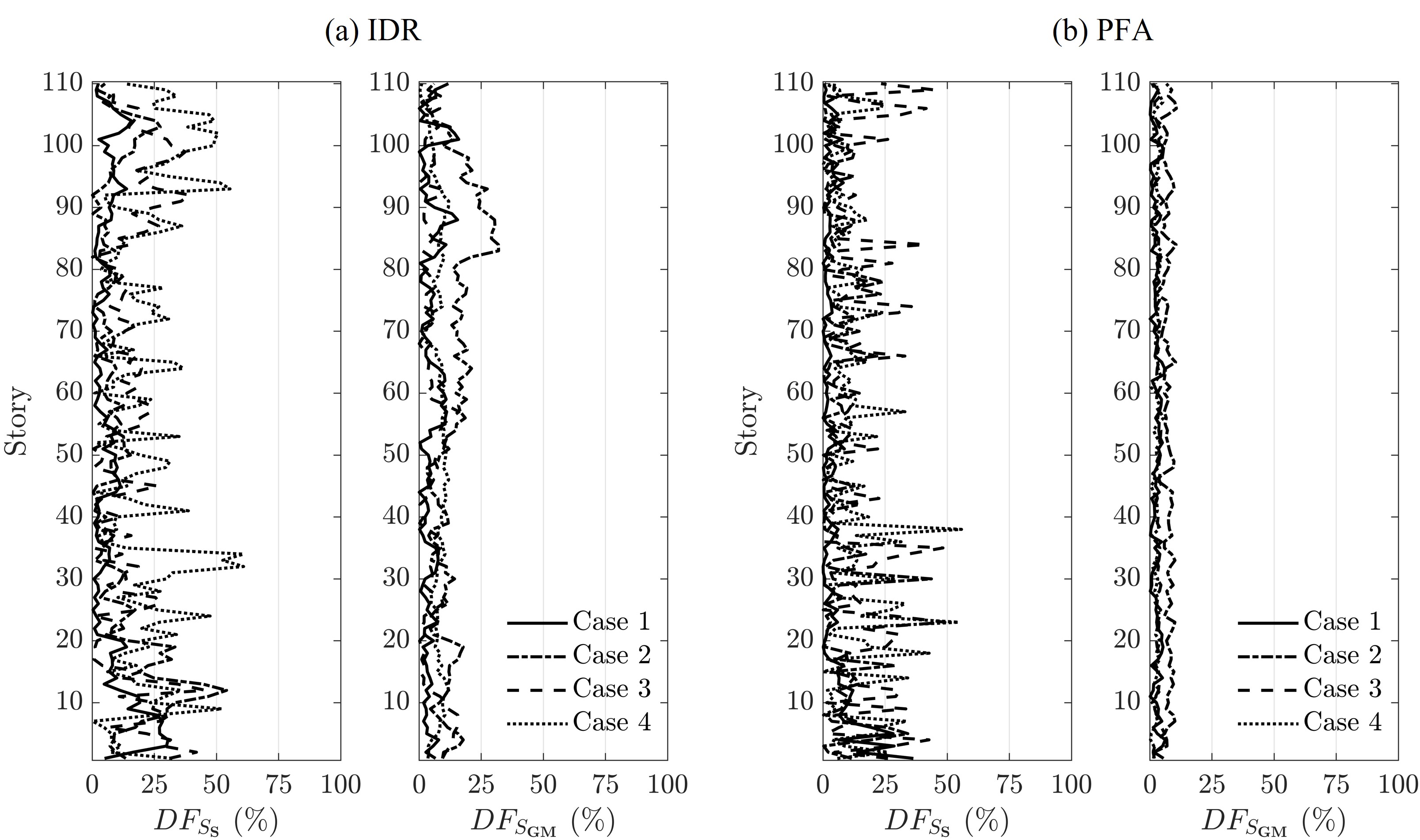}
  \caption{\textbf{Differences in Sobol' sensitivity indices between synthetic and recorded datasets for the long-period tower: (a) IDRs, and (b) PFAs}. Differences are evaluated under the four structural parameter distribution cases.}
  \label{Fig_MDOFtall_DF_S}
\end{figure}

For the long-period tower, sensitivity indices and their discrepancies, shown in Figures~\ref{Fig_MDOFtall_S} and~\ref{Fig_MDOFtall_DF_S}, exhibit trends similar to those observed for the medium-period building, with overall differences remaining below 30\%. These results confirm the applicability of synthetic ground motions for analyzing global sensitivity, while emphasizing the growing influence of structural uncertainties in taller structures, particularly when the variability in uncertain parameters is large.

\begin{table}[H]
  \caption{\textbf{Sobol' sensitivity indices for IDRs under two groups of input uncertainties for the medium-period building structure}.}
  \label{Tab_MDOF_Su_IDR}
  \centering
  \resizebox{\textwidth}{!}{ 
  \begin{tabular}{c | c >{\columncolor{gray!6}}c c >{\columncolor{gray!6}}c c >{\columncolor{gray!6}}c c >{\columncolor{gray!6}}c | c >{\columncolor{gray!6}}c c >{\columncolor{gray!6}}c c >{\columncolor{gray!6}}c c >{\columncolor{gray!6}}c}
    \toprule
    \multirow{3}{*}{Story} &     \multicolumn{8}{c|}{Sensitivity index under recorded motions (\%)} & \multicolumn{8}{c}{Sensitivity index under synthetic motions (\%)} \\
    \cmidrule(lr){2-9} \cmidrule(lr){10-17} & \multicolumn{2}{c}{Case 1} & \multicolumn{2}{c}{Case 2} & \multicolumn{2}{c}{Case 3} & \multicolumn{2}{c|}{Case 4} & \multicolumn{2}{c}{Case 1} & \multicolumn{2}{c}{Case 2} & \multicolumn{2}{c}{Case 3} & \multicolumn{2}{c}{Case 4} \\
    \cmidrule(lr){2-9} \cmidrule(lr){10-17} & $S_{\mathbf{S}}$ & \cellcolor{white}{$S_{\mathbf{GM}}$} & $S_{\mathbf{S}}$ & \cellcolor{white}{$S_{\mathbf{GM}}$} & $S_{\mathbf{S}}$ & \cellcolor{white}{$S_{\mathbf{GM}}$} & $S_{\mathbf{S}}$ & \cellcolor{white}{$S_{\mathbf{GM}}$} & $S_{\mathbf{S}}$ & \cellcolor{white}{$S_{\mathbf{GM}}$} & $S_{\mathbf{S}}$ & \cellcolor{white}{$S_{\mathbf{GM}}$} & $S_{\mathbf{S}}$ & \cellcolor{white}{$S_{\mathbf{GM}}$} & $S_{\mathbf{S}}$ & \cellcolor{white}{$S_{\mathbf{GM}}$} \\
    \midrule
1 & 13.37 & 59.53 & 10.40 & 66.04 & 5.28 & 84.16 & 0.79 & 87.40 & 16.63 & 55.92 & 12.10 & 71.82 & 4.73 & 80.13 & 0.70 & 91.92 \\
2 & 12.76 & 59.91 & 7.40 & 66.81 & 3.36 & 82.19 & 0.42 & 88.51 & 18.90 & 59.08 & 9.29 & 70.58 & 3.53 & 83.37 & 0.50 & 92.29 \\
3 & 20.05 & 51.64 & 11.91 & 63.60 & 4.71 & 76.21 & 1.25 & 88.43 & 26.21 & 51.74 & 13.93 & 65.15 & 5.43 & 79.04 & 1.66 & 90.73 \\
4 & 28.23 & 42.78 & 18.74 & 55.56 & 10.25 & 69.21 & 3.38 & 86.15 & 34.82 & 42.23 & 21.37 & 56.29 & 11.08 & 72.54 & 3.07 & 86.80 \\
5 & 33.77 & 37.04 & 24.71 & 48.06 & 14.52 & 64.81 & 4.28 & 81.05 & 39.47 & 34.54 & 25.42 & 50.61 & 15.50 & 68.60 & 3.31 & 82.86 \\
6 & 39.02 & 34.98 & 31.11 & 44.94 & 18.79 & 59.71 & 5.39 & 77.47 & 42.58 & 31.64 & 29.10 & 47.33 & 17.66 & 61.77 & 4.76 & 78.48 \\
7 & 43.90 & 32.38 & 36.39 & 41.58 & 21.34 & 55.58 & 7.40 & 77.82 & 44.12 & 30.40 & 31.75 & 48.09 & 20.69 & 61.50 & 5.57 & 76.21 \\
8 & 46.65 & 31.62 & 38.67 & 38.51 & 24.75 & 55.14 & 8.67 & 76.20 & 44.95 & 30.06 & 32.53 & 45.30 & 20.79 & 58.86 & 7.19 & 76.95 \\
9 & 49.02 & 30.15 & 40.63 & 33.41 & 25.59 & 51.63 & 8.13 & 72.67 & 49.82 & 26.87 & 35.39 & 42.21 & 25.76 & 56.58 & 6.30 & 74.20 \\
10 & 53.59 & 26.42 & 42.55 & 29.20 & 28.91 & 48.68 & 10.47 & 72.94 & 54.15 & 24.05 & 40.36 & 38.15 & 29.75 & 52.84 & 8.90 & 70.96 \\
11 & 57.45 & 24.05 & 49.91 & 29.07 & 34.85 & 47.11 & 12.00 & 70.63 & 59.30 & 21.52 & 49.05 & 31.91 & 36.36 & 49.58 & 11.27 & 70.48 \\
12 & 59.77 & 21.57 & 51.72 & 27.88 & 36.94 & 47.11 & 10.87 & 68.37 & 61.19 & 19.23 & 53.89 & 29.91 & 36.75 & 49.34 & 14.25 & 71.59 \\
    \bottomrule
  \end{tabular} 
  }  \\
\end{table}
\begin{table}[H]
  \caption{\textbf{Sobol' sensitivity indices for PFAs under two groups of input uncertainties for the medium-period building structure}.}
  \label{Tab_MDOF_Su_PFA}
  \centering
  \resizebox{\textwidth}{!}{ 
  \begin{tabular}{c | c >{\columncolor{gray!6}}c c >{\columncolor{gray!6}}c c >{\columncolor{gray!6}}c c >{\columncolor{gray!6}}c | c >{\columncolor{gray!6}}c c >{\columncolor{gray!6}}c c >{\columncolor{gray!6}}c c >{\columncolor{gray!6}}c}
    \toprule
    \multirow{3}{*}{Story} &     \multicolumn{8}{c|}{Sensitivity index under recorded motions (\%)} & \multicolumn{8}{c}{Sensitivity index under synthetic motions (\%)} \\
    \cmidrule(lr){2-9} \cmidrule(lr){10-17} & \multicolumn{2}{c}{Case 1} & \multicolumn{2}{c}{Case 2} & \multicolumn{2}{c}{Case 3} & \multicolumn{2}{c|}{Case 4} & \multicolumn{2}{c}{Case 1} & \multicolumn{2}{c}{Case 2} & \multicolumn{2}{c}{Case 3} & \multicolumn{2}{c}{Case 4} \\
    \cmidrule(lr){2-9} \cmidrule(lr){10-17} & $S_{\mathbf{S}}$ & \cellcolor{white}{$S_{\mathbf{GM}}$} & $S_{\mathbf{S}}$ & \cellcolor{white}{$S_{\mathbf{GM}}$} & $S_{\mathbf{S}}$ & \cellcolor{white}{$S_{\mathbf{GM}}$} & $S_{\mathbf{S}}$ & \cellcolor{white}{$S_{\mathbf{GM}}$} & $S_{\mathbf{S}}$ & \cellcolor{white}{$S_{\mathbf{GM}}$} & $S_{\mathbf{S}}$ & \cellcolor{white}{$S_{\mathbf{GM}}$} & $S_{\mathbf{S}}$ & \cellcolor{white}{$S_{\mathbf{GM}}$} & $S_{\mathbf{S}}$ & \cellcolor{white}{$S_{\mathbf{GM}}$} \\
    \midrule
1 & 5.30 & 74.29 & 5.32 & 74.99 & 1.82 & 78.28 & 0.71 & 83.77 & 5.51 & 72.17 & 4.70 & 81.13 & 2.15 & 84.16 & 1.00 & 83.61 \\
2 & 5.55 & 77.09 & 4.08 & 76.40 & 2.13 & 81.33 & 0.76 & 82.77 & 6.79 & 72.88 & 5.32 & 78.13 & 2.99 & 83.52 & 0.58 & 83.87 \\
3 & 7.90 & 79.53 & 3.66 & 77.63 & 3.38 & 80.11 & 1.14 & 83.93 & 7.05 & 73.11 & 5.29 & 80.76 & 3.50 & 83.90 & 1.36 & 85.83 \\
4 & 7.49 & 78.85 & 4.58 & 77.10 & 3.47 & 81.91 & 0.76 & 85.98 & 7.05 & 73.56 & 5.27 & 80.64 & 3.38 & 83.84 & 1.02 & 88.01 \\
5 & 6.79 & 78.11 & 4.17 & 76.76 & 3.75 & 83.87 & 1.42 & 87.66 & 8.58 & 74.57 & 5.79 & 80.72 & 4.81 & 85.27 & 1.20 & 88.76 \\
6 & 8.15 & 80.51 & 5.95 & 78.10 & 2.80 & 84.27 & 1.15 & 86.63 & 9.26 & 73.73 & 7.43 & 82.95 & 3.84 & 85.91 & 1.01 & 88.25 \\
7 & 9.07 & 79.03 & 6.82 & 77.58 & 3.88 & 82.56 & 1.40 & 86.35 & 8.10 & 71.52 & 5.81 & 81.87 & 3.16 & 83.59 & 2.16 & 86.35 \\
8 & 8.49 & 78.37 & 5.48 & 77.26 & 1.03 & 82.02 & 1.07 & 86.55 & 7.82 & 72.51 & 4.43 & 81.60 & 1.63 & 84.24 & 1.82 & 88.18 \\
9 & 10.78 & 76.22 & 6.68 & 74.56 & 3.06 & 81.31 & 0.74 & 89.01 & 9.88 & 71.77 & 5.58 & 78.84 & 4.73 & 86.75 & 1.19 & 87.28 \\
10 & 11.61 & 71.86 & 11.12 & 72.31 & 5.42 & 80.75 & 0.99 & 86.83 & 12.52 & 67.72 & 8.97 & 78.75 & 4.67 & 83.82 & 1.53 & 86.00 \\
11 & 16.85 & 69.12 & 12.94 & 67.74 & 7.07 & 76.57 & 0.90 & 86.82 & 18.10 & 62.89 & 11.56 & 71.83 & 8.00 & 79.32 & 1.12 & 87.11 \\
12 & 14.39 & 66.08 & 12.34 & 66.52 & 7.20 & 76.43 & 1.04 & 84.22 & 16.48 & 62.38 & 11.27 & 71.01 & 9.28 & 78.08 & 1.68 & 85.91 \\
    \bottomrule
  \end{tabular} 
  }  \\
\end{table}

\section{Synthesis of the comparative analysis} \label{Summary}

\begin{figure}[H]
  \centering
  \includegraphics[scale=0.60] {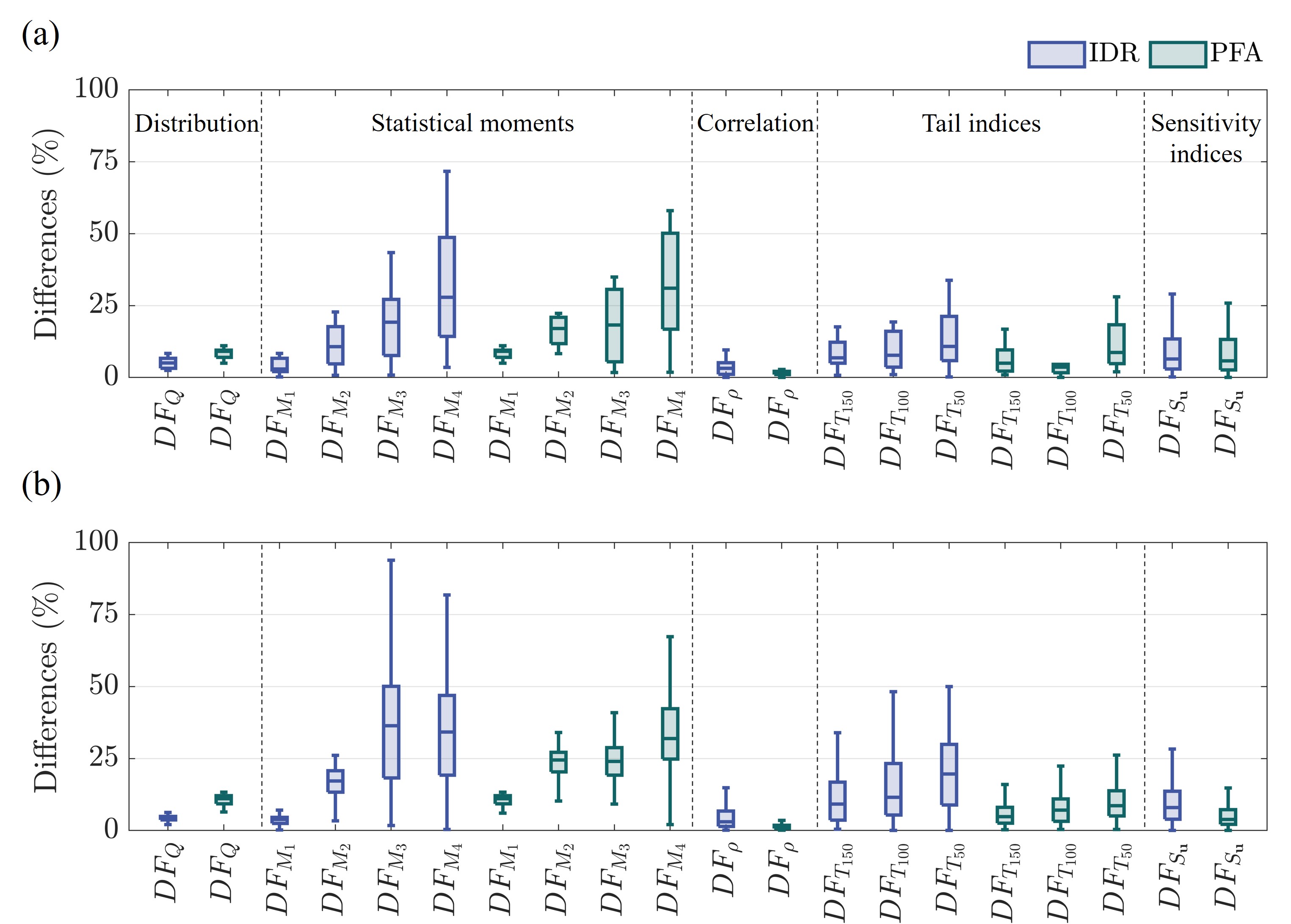}
  \caption{\textbf{Summary of differences in five seismic response metrics under synthetic and recorded motions for (a) the medium-period building and (b) the long-period tower}.}
  \label{Fig_DF_summary}
\end{figure}

\noindent The comparative analysis of stochastic seismic responses—quantile-based distribution differences ($DF_Q$), statistical moment differences ($DF_{M_k}$), correlation differences ($DF_{\rho}$), tail index differences ($DF_{T_k}$), and sensitivity index differences ($DF_{S_\mathbf{u}}$)—is synthesized in Figure~\ref{Fig_DF_summary}, which presents box plots summarizing differences across all stories of the medium-period building and long-period tower. For metrics such as $DF_{M_k}$ and $DF_{T_k}$, individual components (e.g., specific moments and tail exceedance levels) are further distinguished.

The findings reveal several dominant trends, summarized as follows:
\begin{itemize}
\item \textbf{Consistency in global response characteristics}: Synthetic and recorded ground motions exhibit close agreement in global response metrics, including distributions, lower-order moments (mean and variance), inter-EDP correlations, and sensitivity indices. Differences in these metrics generally fall within 10\%–20\%, confirming the effectiveness of synthetic motions in capturing standard seismic response behavior. Moreover, synthetic motions reliably preserve global trends in sensitivity rankings and correlation patterns across EDPs, supporting their utility for routine seismic analysis under well-defined earthquake scenarios. This robustness highlights the value of synthetic motions for efficiently representing central tendencies and variability in seismic responses.
    
\item \textbf{Significant discrepancies in extreme behaviors}: Metrics associated with extreme behaviors, such as tail indices and higher-order normalized moments (skewness and kurtosis), exhibit pronounced discrepancies, often exceeding 50\%. These differences are particularly prominent for story drift responses in long-period structures. Synthetic motions consistently underpredict the frequency and severity of large excursions, a limitation that cannot be identified by examining only low-order statistics. Most previous studies have validated synthetic motions primarily using spectral compatibility or mean/variance agreement, which overlook these tail discrepancies. This study reveals that such underestimation of extremes poses non-conservative risks, particularly in performance-based seismic design, where rare events govern decision-making. These results highlight that synthetic ground motions—while useful for average behavior—may not be suitable for simulating extreme responses without further refinement.
    
\item \textbf{Amplification of discrepancies in long-period structures}: Discrepancies are particularly pronounced in the long-period tower due to its heightened sensitivity to low-frequency ground motion components. For example, IDRs in the long-period tower exhibit larger differences in both higher-order moments and tail indices compared to the medium-period building. Recorded motions, especially those from near-fault events, contain intricate features that amplify extreme responses in long-period structures. In contrast, synthetic motions, constrained by Gaussian-like assumptions, often fail to reproduce these features, resulting in smoothed tail behavior. These findings underscore the critical role of recorded motions in accurately assessing seismic demands for tall buildings and other long-period structures.

\item \textbf{Challenges in ground motion representation}: The observed differences in seismic responses between recorded and synthetic motions can be traced to three primary sources. First, ground motion selection is limited by the size of candidate motion pools, while generation methods must balance computational efficiency with spectral matching accuracy. Second, a single spectral acceleration spectrum can correspond to multiple ground motion realizations, introducing variability in time-domain characteristics. Third, these factors collectively amplify discrepancies in higher-order moments and tail metrics, as these measures are particularly sensitive to the sparse data in the tails of response distributions. These challenges underscore the complexity of mapping spectral acceleration uncertainty to ground motion variability.
    
    \item \textbf{Relative importance of aleatory and epistemic uncertainties}:  
    Sensitivity analysis shows that ground motion variability  consistently dominates structural parameter variability  in driving seismic responses. However, for both structural models, the relative influence of structural uncertainties increases with building height, particularly for IDRs in upper stories, where structural variability can surpass ground motion variability in some cases. This effect is most pronounced when structural variability is high but remains notable across all studied cases with varying distribution types. In contrast, for PFAs, ground motion variability remains the dominant factor across all cases, regardless of structural variability. These sensitivity patterns are consistent across both synthetic and recorded datasets, confirming the applicability of synthetic motions for capturing global sensitivity trends, while emphasizing the growing influence of structural uncertainties in taller structures with large parameter variability.

\end{itemize}

\section{Conclusions} \label{Conclusion}

This study presents a detailed comparative analysis of stochastic seismic responses under synthetic and recorded ground motions, focusing on five key metrics: probability distributions, statistical moments, correlations, tail indices, and sensitivity indices. Both ground motion datasets are calibrated to a shared target spectrum, ensuring consistency in spectral characteristics and enabling a robust comparison of their effects on structural responses. Two structural models—a 12-story medium-period building and a 110-story long-period tower—are used to investigate how these response metrics vary with ground motion type across different dynamic regimes.

The results show that synthetic motions are reliable surrogates for recorded motions in capturing global response trends. Across both structural models, differences in distributions, mean and variance, correlations, and variance-based sensitivity indices are generally small (within 10\%–20\%), validating their suitability for routine seismic design and global probabilistic response modeling. These findings underscore the value of synthetic motions as efficient tools for representing central tendencies and overall variability in seismic responses. However, significant discrepancies arise in metrics related to extreme behaviors, particularly for long-period structures. Metrics associated with extreme events, including the tails of story drift distributions and higher-order moments, exhibit large discrepancies, often exceeding 50\%. These discrepancies highlight the limitations of synthetic motions in replicating the complex, non-Gaussian features of recorded motions, such as pulse-like effects and extreme variability—limitations that are critical for rare-event analysis and the design of tall buildings or other long-period structures. This underscores the need for caution when using synthetic motions in rare-event risk assessments or design of tall buildings.

In conclusion, when carefully calibrated, synthetic motions provide a computationally efficient and statistically consistent means of capturing global seismic response characteristics, but their use remains limited for applications requiring accurate representation of rare events or long-period structural responses. This distinction is vital: relying solely on synthetic motions without addressing tail behavior may compromise the reliability of rare-event analyses and performance-based seismic design.

Future research could focus on enhancing stochastic ground motion models to better replicate non-Gaussian features and complex earthquake phenomena, bridging the gap between computational efficiency and real-world accuracy in seismic design and risk assessment. In addition, the structural models used in this study are idealized shear-type MDOF systems with uniform story stiffness and bilinear lateral force–displacement behavior. While this idealization enables controlled comparisons, it excludes element-level inelasticity, $P$–$\Delta$ effects, and vertical irregularities. Future extensions may incorporate more detailed nonlinear modeling and selected response metrics to improve the applicability of the findings to performance-based seismic design. Moreover, synthetic ground motions are generated using an evolutionary spectrum–based procedure, which is calibrated to reproduce a prescribed unconditional target response spectrum, including its median, dispersion, and period-wise correlation structure. The comparative conclusions are therefore interpreted primarily in relation to these target spectral statistics rather than the specific implementation details of the chosen ground motion generator. Provided that an alternative spectrum-compatible method can achieve a comparable match to the same target model, the main qualitative findings are expected to remain consistent. A systematic evaluation of different spectrum-compatible generation methods constitutes a promising direction for future research.

\section*{Acknowledgments}

This research was supported by the Pacific Earthquake Engineering Research Center (PEER) under grant NCTRZW. Additional support was provided by the Basic Science Research Program through the National Research Foundation of Korea (NRF), funded by the Ministry of Education (RS-2024-00407901). The authors gratefully acknowledge these supports.

\section*{Data availability}
The source codes are available for download at \url{https://github.com/Jungh0Kim/GM_Comparative}.

\appendix
\renewcommand{\theequation}{A.\arabic{equation}}
\renewcommand{\thefigure}{A.\arabic{figure}}
\renewcommand{\thetable}{A.\arabic{table}}
\setcounter{figure}{0} 
\setcounter{table}{0} 

\section{Algorithms for ground motion selection and generation} \label{App:GM_algo}

\noindent This study employs recorded and synthetic ground motion datasets that are calibrated to match a shared target spectrum, ensuring consistency in spectral median, variance, and correlation structure. The datasets are constructed using two advanced algorithms: a ground motion selection method \cite{baker2018improved} for recorded motions and a ground motion generation method \cite{yanni2024probabilistic} for synthetic motions. The procedures are as follows, with both methods sharing the first two steps:

\begin{itemize}
    \item \textbf{Define the target spectrum:} Specify a target response spectrum Sa($T$), including its median, variance, and correlation structure. The target spectrum can be derived from a design spectrum or a GMPE. The natural logarithm of Sa($T$) is modeled as a Gaussian process indexed by the period $T$.

    \item \textbf{Generate target realizations:} Create $N_t$ realizations of the target spectrum, $\{\hat{S}a^i(T)\}_{i=1}^{N_t}$, to guide the selection or generation of ground motions.

    \item \textbf{Obtain the ground motion dataset:}
    \begin{enumerate}
    \item[a)] \textbf{For recorded motions}: The selection algorithm described in \cite{baker2018improved} is employed. This approach selects $N_t$ ground motions from a candidate dataset based on their similarity to the target spectrum. The process involves computing error metrics between the candidate motion spectra and each target spectrum realization, selecting those with the smallest error, and optionally scaling their amplitudes for better spectral matching.
    
    \item[b)] \textbf{For synthetic motions}: The generation algorithm described in \cite{yanni2024probabilistic} is used. This method employs spectral representation techniques \cite{vanmarcke1977simulated} with time-frequency modulating function to generate artificial ground motions. Iterative adjustments are made to align the generated motions with the target spectrum. A baseline correction ensures realistic time histories, resulting in non-stationary accelerograms compatible with the target spectrum.  
    \end{enumerate}
\end{itemize}

These algorithms ensure that both the recorded and synthetic datasets closely align with the specified target spectrum, facilitating robust comparisons of their impacts on seismic response metrics.

\renewcommand{\theequation}{B.\arabic{equation}}
\renewcommand{\thefigure}{B.\arabic{figure}}
\renewcommand{\thetable}{B.\arabic{table}}
\setcounter{figure}{0} 
\setcounter{table}{0} 

\section{Computation of sensitivity indices using Saltelli's method} \label{App:Saltelli}

\noindent Variance-based sensitivity indices are critical for quantifying the contributions of input uncertainties to seismic responses. However, their computation can be computationally intensive, particularly for high-dimensional models frequently encountered in seismic response analyses. Traditional variance decomposition methods often require a double-loop integration process, leading to a substantial number of model evaluations. To address this challenge, Saltelli’s scheme \cite{saltelli2010variance} is utilized in this study. This method employs stratified quasi-random sampling to compute sensitivity indices efficiently, significantly reducing computational costs, and has been widely applied in engineering domains \cite{shin2013addressing,mahmoudi2019global,dela2022multi}.

The approach leverages two quasi-random sampling matrices, $\vect{X}_{\mathbf{A}}$ and $\vect{X}_{\mathbf{B}}$, each of size $N \times n$, where $N$ denotes the number of samples and $n$ represents the input dimension. Corresponding output matrices, $\vect{Y}_{\mathbf{A}}$ and $\vect{Y}_{\mathbf{B}}$, are of size $N \times m$, where $m$ is the output dimension. The Sobol' sensitivity index for the $k$-th output, $Y_k$, with respect to an input group $\vect{X}_{\mathbf{u}}$, is computed as follows:
\begin{equation}  \label{Eq:SaltelliFirst}
S_\mathbf{u}^k = \frac{\frac{1}{N}\sum_{j=1}^{N} Y_{\mathbf{B},k}^{(j)} \left(Y_{\mathbf{AB_\mathbf{u}},k}^{(j)} - Y_{\mathbf{A},k}^{(j)} \right)}{\frac{1}{N}\sum_{j=1}^{N} \left(Y_{\mathbf{A},k}^{(j)} \right)^2 - \left(\frac{1}{N}\sum_{j=1}^{N} Y_{\mathbf{A},k}^{(j)} \right)^2} \,.
\end{equation}
Here, $Y_{\mathbf{A},k}^{(j)}$ denotes the $j$-th row of $\vect{Y}_{\mathbf{A},k}$, and $\vect{Y}_{\mathbf{AB_\mathbf{u}}}$ represents the output matrix corresponding to the combined input matrix $\vect{X}_{\mathbf{AB_\mathbf{u}}}$. This combined input matrix, $\vect{X}_{\mathbf{AB_\mathbf{u}}}$, is constructed by replacing the columns of $\vect{X}_{\mathbf{A}}$ corresponding to the variables in $\mathbf{u}$ with those from $\vect{X}_{\mathbf{B}}$. This process facilitates the efficient computation of sensitivity indices while preserving the stratification of quasi-random samples. The computational cost of the method is $(n_G + 2)N$, where $n_G$ represents the number of input groups. This efficiency makes the method suitable for seismic sensitivity analysis, which often involves computationally demanding NLRHAs.

\bibliography{Comparative_Seismic}


\end{document}